\documentclass[preprint,prb,draft,showpacs,12pt]{revtex4}
\usepackage{amssymb,latexsym,,epsf,colordvi,graphicx,wrapfig,rotate,epsf}
\usepackage[usenames]{color}

\newcommand{\comment}[1]{}

\newcommand{\n}{\nonumber}
\newcommand{\x}{{\rm x}}
\newcommand{\y}{{\rm y}}
\newcommand{\z}{{\rm z}}
\newcommand{\vi}{{\rm vib}}
\newcommand{\be}{\begin{equation}}
\newcommand{\ee}{\end{equation}}
\newcommand{\bea}{\begin{eqnarray}}
\newcommand{\eea}{\end{eqnarray}}
\newcommand{\ket}[1]{\left|#1\right\rangle}
\newcommand{\bra}[1]{\left\langle #1\right|}

\renewcommand{\(}{\left(}
\renewcommand{\)}{\right)}
\renewcommand{\[}{\left[}
\renewcommand{\]}{\right]}

\begin{document}

\title{Electron-nuclear correlations for photo-induced dynamics in molecular dimers}

\author{Dmitri S. Kilin, 
        Yuri V. Pereversev,
    and Oleg V. Prezhdo
\footnote{Corresponding author. E-mail: prezhdo@u.washington.edu.}
}

\affiliation{University of Washington, Seattle WA, 98195\\
{\rm Submitted to J. Chem. Phys. \today}}


\begin{abstract}

Ultrafast photoinduced dynamics 
of electronic excitation in molecular dimers 
is drastically affected by the dynamic reorganization 
of inter- and intra- molecular nuclear configuration 
modeled by a quantized nuclear degree of freedom 
[Cina et. al, J. Chem Phys. {\bf 118}, 46 (2003)]. 
The dynamics of the electronic population and nuclear coherence
is analyzed by solving 
the chain of coupled 
differential equations for 
population inversion, electron-vibrational correlation, etc. 
[Prezhdo, Pereverzev, J. Chem. Phys. {\bf 113} 6557 (2000)]. 
Intriguing results are obtained in the approximation 
of a small change of the nuclear equilibrium 
upon photoexcitation.
In the limiting case of resonance 
between the electronic energy gap
and the frequency of the nuclear mode 
these results are justified 
by comparison to the exactly solvable 
Jaynes-Cummings model. 
It is found that the photoinduced processes 
in the model dimer are arranged according 
to their time scales: 
(i) fast scale of nuclear motion, 
(ii) intermediate scale of dynamical redistribution 
of electronic population between excited states 
as well as growth and dynamics 
of electron-nuclear correlation, 
(iii) slow scale of electronic population 
approach to the quasi-equilibrium distribution, 
decay of electron-nuclear correlation, 
and decrease of the amplitude 
of mean coordinate oscillation. The latter processes are
accompanied by a noticeable growth of the 
nuclear  coordinate dispersion
associated with the overall nuclear wavepacket width. 
The demonstrated quantum relaxation features 
of the photoinduced vibronic dynamics
in molecular dimers are obtained by a simple method, 
applicable to systems with many degrees of freedom.
\end{abstract}

\maketitle

\section{Introduction}
	Recent achievements of femtosecond spectroscopy~\cite{Zewail} open new windows into the 
world of electron and exciton transfer dynamics in molecular systems, which play 
dominant roles in a variety of problems in physics, technology, atmospheric 
photochemistry, and biology~\cite{brook03,stier03,balzani_scandola,Rienk_van_Grondelle-BOOK,monat02}.
Electron and exciton transfer drive formation and breaking of 
chemical bonds, corrosion reactions, ion tunneling microscopy, enzymatic activity in 
living cells, energy storage by adenosine triphosphate and harvesting of solar energy. For 
instance, the harvesting and storage of solar energy in the natural photosynthetic 
complexes, such as bacteriochlorophyll~\cite{nobel_prize88} or Rhodobacter sphaeroids~\cite{balzani_scandola}, 
rely on the specific alignments of the light-harvesting units that favor efficient light absorption and 
subsequent energy transfer. 
The natural and artificial light harvesting complexes (LHC)~\cite{marin02}  
 involve exciton transfer of the primary photo-excitation in the antennae complex~\cite{Grahm_Fleming_AND_Xantippe}, 
followed by a subsequent electron transfer that takes place in the reaction center of the 
natural LHC and porphyrin arrays of artificial LHC~\cite{wasi92,mccusker00}.

	The dynamics of electron and energy transfer, and formation of entangled 
electron-phonon states are investigated by a wide arsenal of femtosecond experimental 
techniques, which by application of one or several short laser pulses, provide time 
resolved information on couplings, correlations and states 
of molecular aggregates~\cite{Zare,Hochstrasser,textbook-spectroscopy,Tannor-BOOK,non-lin-optics,tretiak_review}.
The multi-pulse time resolved techniques cover various frequency ranges, including X-ray, 
visible and infrared, and serve for determination of population, location and 
phase of electron-nuclear states of molecular systems. Thus, a recently developed Raman 
X-ray spectroscopy follows electronic state dynamics~\cite{Takayoshi_ask_Piryatinski}. 
Laser pulses in the infrared diapason are used to investigate populations and mutual 
correlations of vibrational modes~\cite{Fayer,Tokmakoff}.
Visible frequency short laser pulses are applied to measure electronic state populations and state-to-state correlations~\cite{Jonas}.
Positions of nuclear wavepackets are detected by pump-probe spectroscopic techniques~\cite{Zewail}. Relative 
phases of electron-nuclear wavepackets of molecular aggregates are obtained using the 
nonlinear wavepacket interferometry~\cite{sher91,sher92,Cina,Shapiro}.

	Theoretical methods describing exciton and electron transfer dynamics appeal, in 
one way or another, to both electronic and nuclear degrees of freedom~\cite{marc56,foerster,stepanov} and most 
commonly in chemical physics employ the notion of a reaction coordinate~\cite{kuhn_may}. 
The reaction 
coordinate quantifies changes in the electronic states due to nuclear mode dynamics and 
is based on the fact that transfer is most effective for specific configurations of the 
nuclear subsystem. These nuclear configurations correspond to crossings of the 
potential surfaces and match the Franck-Condon window for the source and target states. 
In many cases, particularly for large systems, the nuclear subsystem leaves and never 
returns to the optimal configuration region due to stochastic noise and destructive 
interference among nuclear modes of different frequencies involved in the dynamics. 
This dynamic mismatch and dephasing of the nuclear modes is responsible for relaxation and decoherence~\cite{ross97} 
in the exciton and electron transfer dynamics.


	The coupled electron-phonon dynamics can be described by a number of 
theoretical approaches developed in chemical physics~\cite{schatz-ratner-book}, 
including the Gaussian wave-packet approaches~\cite{Heller70,Mukamel89}, 
the semiclassical approximations~\cite{bill_miller}  based on Feynman path integrals~\cite{feynman}, 
and non-adiabatic molecular dynamics~\cite{coke93,tull98,schw03}. 
The quantum-classical mean-field approach and 
its multiconfiguration mean-field and surface hopping extensions are applied to a wide 
range of problems, including gas-phase scattering, surface~\cite{stier03}, solution and biological 
chemistry~\cite{referees}. Semiempirical approaches introduce quantum corrections to classical 
mechanics. The recently developed Quantized Hamilton Dynamics (QHD) method~\cite{ross97,pere00,Prezhdo2002,pere02,pahl02,broo03} 
offers a reduced description of quantum nuclear motion by 
disentangling coordinate, momentum, dispersion and higher order variables for each 
nuclear mode. A many body closure~\cite{Wick1950} terminates the infinite chain of coupled equations 
reducing the number of equations and saving the calculation time. While the 
semiclassical and mean-field approaches are very effective from the numerical point of view, 
they provide approximate solutions. In particular, the mean-field approaches typically cannot 
describe the dynamic creation of superposition and entanglement between states leading 
to branching of nuclear dynamics. 


Theoretical models that are able to describe the superposition and branching 
feature of coupled electron-phonon dynamics are available in the solid-state theory~\cite{Leggett} 
and  quantum optics theories of atom-field interactions~\cite{general_QO}.  
The quantum optics approaches to the 
description of photon modes coupled to atomic quantum states~\cite{Allen_Eberly} 
can be adopted to study 
electron-phonon states in chemical, biological and nano-systems. A simple eigenstate 
solution for the correlated dynamics of two electronic states of an atom (fermion degree 
of freedom) coupled to a laser mode (boson degree of freedom), presented in 1963 by 
Jaynes and Cummings~\cite{Jaynes-Cummings63,KnightJCM}, 
takes an important place among the optical methods. 
The Jaynes-Cummings solution was further developed by the operator algebra methods~\cite{modernALGEBRAjcm}. 
Known are the operator solutions for the lowering and raising operators~\cite{Ackerhart}
 and for the electronic state population~\cite{Scully_BOOK} that shows collapses~\cite{cumi65} and revivals~\cite{naro80}. 
Interesting expressions are found for the 
uncertainty relationships for the operators describing the two-level atom interacting with 
a photon field~\cite{Berman_Gena_BOOK}. The generalized operator solutions lead to a well-developed formalism in 
the Fokker-Planck equation formulation~\cite{Carmichael_BOOK}. The collapses and revivals in the two-level atom 
excitation dynamics of the Jaynes-Cummings model represent the simplest case of 
quantum relaxation common for exciton and electron transfer in molecular complexes~\cite{redf55,kilin2000,poll94}.


The quantum optics and QHD approaches are combined in this paper for the 
description of molecular aggregates, whose photoexcitation dynamics is strongly affected 
by dynamic rearrangements of vibrational degrees of freedom. The combination of 
methods is very effective for the calculation of the relaxation behavior of exciton transfer 
in molecular aggregates.  Extending the original QHD approach that focuses onto 
semiclassical dynamics of nuclear modes, a QHD approximation is developed in present for the 
dynamics of the coupled electronic and nuclear variables, concentrating on the difference 
of the electronic state populations, i.e., population inversion. The resulting equations are 
very simple to be applied to large condensed phase chemical systems. Without reference 
to a thermal bath, as typical in the quantum relaxation theory, the QHD approximation for 
the dynamics of the Jaynes-Cummings model shows quantum relaxation features. The 
relaxation character of the exciton and electron transfer dynamics arises naturally in our 
approach due to destructive interference of quantum states, technically similar to 
quantum beats. The complex dynamics of the electron population in the 
Jaynes-Cummings model is made more understandable by a time-scale hierarchy \cite{kenkre95,S_Ya_Kilin_review_1987} of the 
relevant dynamical processes.

The paper is organized as follows. 
The relevant theoretical tools are introduced in Section~\ref{Model}. 
The calculated quantum dynamics are analyzed in Section~\ref{Analysis_of_dynamics}, including a detailed discussion of the vibronic wavepacket. 
Section~\ref{Analysis_of_uncertainty} considers approach of the system to a state with large uncertainty. 
The physical processes in the system are organized in accordance with their characteristic times in Section~\ref{Time_scales}. 
The scenarios of experiments that can be described by the current method, and where the dynamical 
features found in this paper can play a dominant role, are discussed in Section~\ref{Discussion}. 
Finally, Section~\ref{Conclusions} summarizes the main conclusions of this work.

 
\section{Model} \label{Model}                                   
Consider dynamics of electronic states in a molecular system.
Typically, the initial populations of the states are derived from a thermal equilibrium.
An external perturbation such as a femtosecond optical pulse quickly, within
$10^{-15}-10^{-14}$~s breaks the equilibrium by enhancing population of excited states.
The non-equilibrium populations evolve in time. The populations of the
excited states are redistributed within an intermediate time interval of a few picoseconds, 
$10^{-12}$s.
The population of the electronic states dephases
due to interaction with inter and intramolecular
vibrational modes.
On a relatively long time scale of nanoseconds, $10^{-9}$s, 
the molecular system is de-excited by spontaneous emission
induced by interaction with zeroth order vacuum oscillations of 
photon modes, or by some other mechanism.
Upon electronic de-excitation, additional electron-vibration dynamics
returns the system to thermal equilibrium. 

The time-dependent processes that occur shortly after pumping are of interest
in the present work.  The focus is on dynamics of a pair of
single-excitation localized
 electronic states $i=0,1$. 
One of the states $i=1$ is pumped initially and is coupled to and
exchanges population with the second state.
The two electronic states will be described in terms of 
the second quantization operators, creating $c_i^+$,
or annihilating $c_i$ the excitation of localized state $i$.
The Hamiltonian describing the picosecond $t\simeq 10^{-12}$s
dynamics of the excited states can be written in terms of the
creation and annihilation operators as
\bea
H_0
    &=&
        \sum_i \epsilon_i c^+_i c_i
      + \sum_i \sum_j J_{ij} c^+_i c_j
\label{EQ_electronic_ham}
\eea
The first term contains the electronic state energy $\epsilon_i$ 
times the
 operator of  number of quanta,
whose expectation value equals one 
$\langle c_i^+c_i\rangle=1$ if state $i$ is populated
and zero if state $i$ is empty.
The second term describes the coupling  between the electronic
states by removing population in state $j$, $c_j$ and
creating population in state $i$, $c^+_i$ with the coupling 
constant $J_{ij}$.
In general, both energy and coupling
depend on nuclear configuration. 
For small fluctuations of the nuclear coordinate $q$
the  dependence of energy and coupling on $q$
can be well represented by the zeroth and first order terms 
in the Taylor series expansion
\bea
\epsilon_i(q)
             &=&
                  \epsilon_i^{(0)}
                + q \frac{\partial \epsilon_i}
                         {\partial q}
                    \left.\right|_{q=q_i}
                + ...,~i=0,1.
\label{EQ_expansion_energy}
\eea
Generally, the expansion coefficients are specific 
for each electronic state.
The Taylor expansions of the coupling constants may also be
different for the forward and backward transitions
\bea
J_{01}
      &=&
            J_{01}^{(0)}
          + q\frac{\partial J_{01}}
                  {\partial q}
                    \left.\right|_{q=q_0} + ... \n \\
J_{10}
      &=&
            J_{10}^{(0)}
          + q\frac{\partial J_{10}}
                  {\partial q}
                    \left.\right|_{q=q_1} + ... .
\label{EQ_expansion_transfer}
\eea
Typically, however, the coupling constants are identical
for both directions
$J_{01}=J_{10}$,
$ {\partial J_{01}}/{\partial q}
 ={\partial J_{10}}/{\partial q}$.
The position independent terms in the expansion of the coupling
constant can be eliminated by diagonalizing of the $q$-independent
part of the electronic Hamiltonian.


The evolution of the nuclear coordinate $q$ augments the 
original electronic Hamiltonian~(\ref{EQ_electronic_ham})
with a vibrational term, which equals $p^2/(2m) + m\omega^2q^2/2$
in the harmonic approximation. The vibrational Hamiltonian
is the same for both electronic states.
Multiplication of the
vibrational Hamiltonian 
            $\left(
                  p^2/(2m) + m \omega^2 q^2/2
             \right)
             \hat {\bf 1}_{\rm el} $ 
by a unit operator in the electronic subspace 
$\hat {\bf 1}_{\rm el} = \sum_i c_i^+c_i$
leads to the following electron-phonon Hamiltonian
\bea
H
  &=&
      \sum_i
            c_i^+c_i
            \left\{
                  \epsilon^{(0)}_i
                + \underbrace{
                              q \cdot
                              \frac{\partial \epsilon_i}
                                   {\partial q}
                              }_{}
                + \frac{p^2}{2m}
                + \underbrace{
                              \frac{m\omega^2q^2}{2}
                              }_{}
            \right\}
     +\sum_i \sum_j
                   c^+_ic_j
                   \left\{
                          q \cdot 
                          \frac{\partial J_{ij}}
                               {\partial q     }
                   \right\}
\label{EQ_total_ham}
\eea
Consider the coordinate-dependent terms indicated in (\ref{EQ_total_ham})
by the underbrace sign in more detail. The equilibrium position $q_i$
for the harmonic motion correlated with the electronic state $i$ can be obtained
explicitly by completing the square:
\bea
\epsilon^{'}_i 
\cdot q 
+ \frac
       {m\omega^2}
       {2}q^2
                  &=&
                      \frac{m\omega^2}{2}
                      \left\{
                             \underbrace{
                                         \epsilon_i^{'} q
                                         \frac{2}{m\omega^2}
                                       + q^2
                                       - \left(
                                               \frac
                                                 {\epsilon^{'}_i} 
                                                 {m\omega^2}
                                         \right)^2}_{}
                          +  \left(
                                  \frac
                                      {\epsilon^{'}_i}
                                      {m\omega^2}
                             \right)^2
                      \right\} \n \\
         &=&
             \frac{m\omega^2}{2}
            \left(
                   q-q_i
           \right)^2
           +  \frac{
                                       \left( \epsilon^{'}_i
                                       \right)^2}
                                      {2m\omega^2},                      
\label{EQ_full_square}
\eea
where $\epsilon_i^{'}\equiv{\partial \epsilon_i}/{\partial q}$,
$
E_i
    =
                  \epsilon_i^{(0)}
               +            
                             {\left(\epsilon^{'}_i\right)^2}
                           / {2m\omega^2}$,
is the renormalized state-specific electronic energy
and $q_i=\epsilon^{'}_i (m\omega^2)^{-1}$. 
The electron-phonon Hamiltonian takes the form
\bea
H
  &=&
           \sum_i
            c_i^+c_i
            \left\{
                    E_i
                  + \frac{m\omega^2}
                         {2}
                    \left(q-q_i\right)^2                              
            \right\}
     +\sum_i \sum_j
                   c^+_ic_j
                   \left\{
                          q\cdot\frac{\partial J}{\partial q}
                   \right\}
\label{EQ_long_el_ph_ham}
\eea
Without loss of generality $q_0=0$, $q_1 \ne 0 $.
The Hamiltonian describing two electronic states 
coupled through a vibrational mode
contains two harmonic potential energy surfaces
corresponding to the two electronic states and
centered at $0$ and $q_1$. The states are coupled 
via a coordinate-dependent term similar to the Holstein Hamiltonian~\cite{hols59}
\bea
H
 &=&
     c_0^+c_0
        \left[
        \underbrace{        
              E_0 + m\omega^2/2 q^2
                   }_{\rm lower~potential}
               + p^2/2m 
       \right]
    +c_1^+c_1        
        \left[
        \underbrace{ 
              E_1
                  + m\omega^2/2 \left(q-q_1\right)^2
                   }_{\rm upper~potential}
               + p^2/2m 
        \right] \n \\
&&    +   \underbrace{
     \left[c_0^+c_1 + c_1^+c_0\right]}_{\rm transfer}
     \cdot q \cdot
     \frac{\partial J }{\partial q} 
\label{EQ_specificly_shifted}
\eea
A two-level system of single excitation states is conveniently represented as
a spin-$1/2$ particle in terms of three spin-projection operators
$\hat S_\x$,
$\hat S_\y$,
$\hat S_\z$, 
which can be expressed using the 
lowering and raising operators 
$\hat S_\x=1/2\left( S_+ + S_- \right)$,
$\hat S_\y=i/2\left( S_+ - S_- \right)$,
$\hat S_\z=1/2\left( S_+S_- - S_-S_+ \right)$.
\bea
S_\pm       &=& S_\x\pm iS_\y, \n \\
S_\pm S_\mp &=& 1/2 \pm S_\z,
\label{EQ_raising_lowering_operators}
\eea
The unique relationships between the spin-$1/2$
raising and lowering operators and the creation 
and annihilation operators of the electronic system
in the language of second quantization are given by
\bea
c_0^+c_0 &\equiv& \ket{0}\bra{0} \equiv S_-S_+, \n \\
c_1^+c_1 &\equiv& \ket{1}\bra{1} \equiv S_+S_-, \n \\
c_1^+c_0 &\equiv& \ket{1}\bra{0} \equiv S_+, \n \\
c_0^+c_1 &\equiv& \ket{0}\bra{1} \equiv S_-. 
\label{EQ_raising_spins}
\eea

The evolution of the vibrational coordinate and momentum
relevant for experiments occurs on an atomic scale
demanding a quantum mechanical treatment.
The quantum
coordinate $q$ and momentum $p$ operators
are conveniently replaced by the harmonic raising $a^+$
and lowering $a$ operators 
\bea
q
  &=&
       \sqrt{\frac{\hbar}{2m\omega}}
            \left( a^+ + a\right), \n \\
p
  &=&
      i
      \sqrt{\frac{\hbar m \omega}{2}}
            \left( a^+ - a \right).
\label{EQ_raising_coordinates}
\eea
The above operator transformations translate the problem
into the second quantization language of raising and lowering
operators for both the electronic and vibrational degrees of freedom.
The relevant dynamics will be represented by product operators
containing spin and harmonic raising and lowering terms.
The product operators describe resonance exchange 
of quanta between vibrational and electronic subsystems.
The shift in the equilibrium coordinates of the
two harmonic potentials is represented in the
second quantization language by the dimensionless parameter 
$\bar \alpha 
           =
              q \sqrt{ {m\omega}/{2\hbar}}
           +i p \sqrt{ {\hbar}/{2m\omega}}$,
related to the original coordinate shift  
\bea
q&=&\sqrt{ \frac{\hbar}{2m\omega} }
     \left( \bar \alpha + \bar \alpha ^*\right).
\label{EQ_coordinate_shift}
\eea
The zero of electronic energy is chosen
so that the initially unoccupied lower state $i=0$
has zero energy.
The physically relevant electronic energy gap
$\hbar \Omega =E_1-E_0$
is independent of energy origin.
With all above changes,
the electron-phonon Hamiltonian reads
\bea
H 
  &=&
        \hbar \omega
      S_-S_+ 
       \left[ 
              a^+a +  1/2 
       \right]
    +  \hbar \omega 
       S_+S_- 
       \left[ 
              (a^++\bar\alpha^*)(a+\bar\alpha) +1/2 
       \right] \n \\
&&  + g\left(  S_+ + S_-   \right)  \left( a^+ +a \right)
    + S_+S_-
        \left[
              \hbar \Omega 
            - \hbar \omega \bar \alpha^* \bar \alpha 
        \right] 
\label{EQ_convenient_ham}
\eea
where $g=\sqrt{{\hbar}/{2m\omega}} \quad {\partial J}/{\partial q}$.

The Hamiltonian~(\ref{EQ_convenient_ham})
represents a dimer 
with a pair of
electronic states 
$\left|0\right>$ and
$\left|1\right>$. 
The states are separated by energy difference 
$\hbar \Omega$ and are coupled 
to a single harmonic vibrational mode.  
The equilibrium positions 
of the potential energy surfaces describing the
vibrational mode in the two electronic states
are shifted with respect to each other by 
$q_1$, specified by the 
dimensionless nuclear reorganization parameter 
$\bar \alpha$, 
as illustrated in 
Fig.~1
This model is 
known as the Marcus model~\cite{marc56}
that has an enormous range of applications to 
exciton, electron, proton transfer
and many other chemical 
reactions~\cite{foerster,kuhn_may,schatz-ratner-book}.

%

In the limit of small nuclear reorganization 
the electron-phonon Hamiltonian transforms 
into a simpler form, known in quantum optics
as the Jaynes-Cummings Hamiltonian~\cite{Jaynes-Cummings63,KnightJCM}. 
\bea
H_{\rm JCM}
           &=&
                {\omega}\(a^+a +1/2\) +   {\Omega}S_+S_-  
             +  {g\(a^+ + a\)}(S_+ + S_-).
\label{EQ-JCM-hamiltonian}
\eea
Here and below $\hbar=1$. 
The Hamiltonian contains three terms corresponding
to the electronic subsystem, vibrational 
subsystem, and excitation transfer.
The approximations below are developed for the Jaynes-Cummings Hamiltonian,
but can be equally obtained for a non-adiabatic Hamiltonian with the
momentum dependent coupling 
${g\(a^+ - a\)}(S_+ + S_-)$
rather than the position dependent coupling 
${g\(a^+ + a\)}(S_+ + S_-)$.
A further simplification of the Jaynes-Cummings 
Hamiltonian is obtained by invoking the rotating
wave or resonance approximation that keeps only those
coupling terms that preserve the total number
of quanta in the combined electron-vibrational system.
\bea
H_{\rm JCM}
           &=&
                {\omega}\(a^+a +1/2\) +   {\Omega}S_+S_-  
             +  {g\(a^+S_- + a S_+ \)}.
\label{EQ-RWA}
\eea
The dynamics that follow from this final form of the Hamiltonian are investigated below.


%

The expectation value of the $S_{\rm z}$ operator is the main focus of the present
study.  This expectation value gives populations of the electronic states
of the dimer and, more generally, is related
to the value of a reaction coordinate in
exciton, electron,
proton transfer and other chemical processes.
In the Heisenberg representation of quantum mechanics, the expectation value of $S_{\rm z}$
evolves in time through the time-dependence of the operator. The wave function
remains fixed and is specified by the initial conditions. The Heisenberg equation
of motion (EOM) for the operator $S_{\rm z}$ is given by its commutator with the
Hamiltonian~(\ref{EQ-RWA}). The time-derivative of $S_{\rm z}$ depends
on other time-dependent operators, whose EOMs are also obtained by commutation 
with~(\ref{EQ-RWA}). 
This leads to an infinite hierarchy of EOMs.
Noting that $S_{\rm z}=1/2\(S_+S_--S_-S_+\)$, the infinite hierarchy can be written
in a compact form using only three additional operators
\bea
          \hat \alpha  &=& a^+S_- + aS_+, \n \\
        \hat \beta     &=& a^+S_- - aS_+, \n \\
       \hat \gamma     &=& a^+a + S_+S_- + \frac 12,
\label{EQ_operators}
\eea
that involve the non-interacting Hamiltonian $\gamma$, 
the interaction part of the Hamiltonian $\alpha$, and 
the auxiliary operator $\beta$. The following
hierarchy of Heisenberg EOMs is obtained
starting with the expectation value of the population inversion operator $S_{\rm z}$ 
\bea
i\frac d{dt}
     \left< 
      \alpha \gamma^n 
     \right> 
            &=& 
                -\delta\left<  \beta \gamma^n \right>, \n \\
i\frac d{dt}
     \left<  
      \beta \gamma^n 
     \right> 
             &=& 
                 -\delta\left< \alpha \gamma^n \right>  
                 +\underbrace{g\left<    S_z \gamma^{n+1} \right>}_{\rm needs~closure}, \n \\
i\frac d{dt}
      \left<    
        S_z \gamma^n 
      \right> 
               &=&
                   g\left<  \beta \gamma^n \right>
\label{EQ_blocks}
\eea
with $n=0,1,2,\dots$.
The detuning $\delta=\Omega-\omega$ 
denotes the difference between the electronic 
energy gap $\Omega$ and 
the vibrational frequency $\omega$. 
The Heisenberg EOMs for the operators are arranged in blocks 
of $n$-th order with three equations per block.
The evolution of the lower order block depends on the evolution of the higher order block 
through the coupling term labeled 
in~(\ref{EQ_blocks}) by underbrace.

Useful approximations to the exact solution 
of the infinite chain of equations~(\ref{EQ_blocks})
can be obtained
by limiting the number of equations to a few lower order blocks
and decomposing the higher order coupling term into a product
of lower order terms in the spirit of the quantized Hamiltonian dynamics (QHD)
approach originally introduced for the classical-like expectation
values of the position and momentum 
operators~\cite{ross97,pere00}.
Decomposition of the higher order expectation values into products
of the lower order ones terminates the infinite hierarchy~(\ref{EQ_blocks})
within a finite number of equations. With no {\it a priori} knowledge
of the relationship between the expectation values of the higher and lower order
operators, 
the contributions
of all possible products of the lower order operators that represent
the higher order operator are taken with equal 
weights~\cite{pere00,pahl02}.
The closure applied to decompose higher order moments of
the position and momentum operators into products of 
the first order
expectation values gives classical Hamiltonian mechanics.
Decomposition of the higher order moments
of position and momentum into products of 
the first and
second moments leads to the well known Gaussian approximation~\cite{Prezhdo2002}.
Generally, the QHD procedure allows one to obtain
simple classical-like EOMs for essentially quantum
mechanical characteristics, such as zero-point energy,
tunneling, quantum state populations, 
state-to-state correlations.

In present, the hierarchy~(\ref{EQ_blocks}) is terminated at $n=1$ requiring
decomposition of the $S_z \gamma^2$ term.  Using the general expression
for the decomposition of a triple product~\cite{pere00}
\bea
\left< \hat A \hat B \hat C \right>
            &\simeq&
                \left< \hat A  \right>\left<  \hat B \hat C \right>
               + \left< \hat B  \right>\left<  \hat A \hat C \right> \n \\
           & & +  \left< \hat C  \right>\left<  \hat A \hat B \right>
              -2\left< \hat A  \right>\left< \hat B  \right>\left< \hat C  \right>
\label{EQ_general_closure}               
\eea
this term is decomposed into
\bea
\left< 
   S_\z \gamma^2 
\right>
          &\simeq&
                   2\left< S_\z \gamma \right>\left< \gamma \right>
                 +  \left< S_\z  \right>\left< \gamma^2 \right>
                 - 2\left< S_\z  \right>\left< \gamma \right>^2,
\label{EQ_JCM_closure}
\eea
yielding a closed system
of six differential equations
\bea
i
\frac{d}
     {dt}
 \left({\begin{array}{c}
\left< \alpha          \right>  \\ 
\left< \beta           \right>  \\ 
\left< S_{\rm z}       \right>  \\ 
\left< \alpha \gamma   \right>  \\
\left< \beta \gamma    \right>  \\ 
\left< S_{\rm z}\gamma \right> 
\end{array}}
\right)
 =
\left({ 
\begin{array}{cccccc}
0        & -\delta  & 0  & 0       & 0         & 0 \\ 
-\delta  & 0        & 0  & 0       & 0         & g \\ 
0        & g        & 0  & 0       & 0         & 0 \\ 
0        & 0        & 0  & 0       & -\delta   & 0 \\ 
0        & 0        & 
                     \left< \gamma ^{2}\right> 
                    -2\left< \gamma \right> ^{2} 
                        & -\delta  & 0         & 2\left< \gamma \right>  \\ 
0        & 0        & 0 & 0        & g         & 0
\end{array}
}\right)
 \left( 
\begin{array}{c}
\left< \alpha          \right>  \\
\left< \beta           \right>  \\
\left< S_{\rm z}       \right>  \\
\left< \alpha \gamma   \right>  \\
\left< \beta \gamma    \right>  \\
\left< S_{\rm z}\gamma \right>
\end{array}
\right). 
\label{EQ_six_equations}
\eea
Usually, closures produce systems of non-linear differential equations, such as the non-linear 
classical Hamiltonian EOMs obtained from the
Heisenberg quantum EOMs by
decomposition of the expectation values of 
higher order moments
of position and momentum operators into products 
of the first order moments~\cite{ross97,pere00}. 
The special properties of 
the Jaynes-Cummings Hamiltonian in the rotating
wave approximation, $\[\hat \gamma, H\]=0$
in particular,
make Eqs.~(\ref{EQ_six_equations})
into a system of linear differential equations.  
The operators $\gamma$ and $\gamma^2$ 
appearing in the closure~(\ref{EQ_JCM_closure}) are integrals of
motion of the Hamiltonian~(\ref{EQ-RWA}). 
The quadratic $\left<\gamma\right>^2$ term in~(\ref{EQ_six_equations}) is
a constant specified by the initial conditions.
Equations~(\ref{EQ_six_equations}) give a very simple approximation
to the dynamics of the electron-phonon system.


\section{Analysis of dynamics} \label{Analysis_of_dynamics}                   
%

The closed system of Heisenberg equations~(\ref{EQ_six_equations}) can be solved
analytically for the expectation value of the population inversion $\left< S_\z \right>$.
The time-evolution of $\left< S_\z \right>$ is given by a superposition of two cosine functions,
\bea
\left<S_\z \right>(t) 
     = \left<S_\z \right>(t=0) &+& \frac
                          {g^2\sqrt{\gamma_0}\left[\sqrt{\gamma_0}+1\right]}
                          {2\omega_1^2}
                                       \(1-\cos\omega_1 t\) \n \\
                 &+& \frac
                          {g^2\sqrt{\gamma_0}\left[\sqrt{\gamma_0}-1\right]}
                          {2\omega_2^2}
                                       \(1-\cos\omega_2 t\), 
\label{EQ_solution}
\eea 
 whose frequencies are defined by
\bea
\omega_1^2   &=&  {\delta^2}+ {2g^2}\(\gamma_0 + \sqrt{\gamma_0}\), \n \\
\omega_2^2   &=&  {\delta^2}+ {2g^2}\(\gamma_0 - \sqrt{\gamma_0}\), \n \\
\gamma_0     &=& \left< \gamma \right> (t=0).            
\label{EQ_frequencies}
\eea
The difference in the squares of the frequencies is given by the product
of the square of the electron-phonon coupling constant $g^2$ and the 
square root of the
mean number of 
quanta in the 
system
$\sqrt{\gamma_0}$. As should be expected,
a perfect oscillation in the electronic state population is observed with zero
electron-phonon coupling. As the strength of the electron-phonon coupling grows,
the difference between the two frequencies increases, leading to the dephasing of 
the electronic state population and a corresponding decrease in the oscillation amplitude, 
Fig.~2.
The oscillation of population inversion will proceed more rapidly 
if the vibrational mode contains more energy.
Vibrational motion of a large amplitude
accelerates transfer of electronic population.
The time evolution of the coupled 
electron-phonon system is illustrated
in Fig.~2
for
the initial conditions
\bea
 \left.  \left<  S_\z  \right>   \right| _{t=0} &=&-\frac{1}{2}, \n \\
 \left. \left< \gamma \right>   \right| _{t=0} &=&<a^{+}a>    , \n \\
 \left. \left< \alpha \right>   \right| _{t=0} &=&0           , 
 \left. \left< \beta \right>    \right| _{t=0} =0           .
\label{EQ_initial_conditions}
\eea
The vibrational mode
is prepared in a quasiclassical coherent state
described in the coordinate representation
by a Gaussian displaced from the vibrational 
equilibrium.
All electronic population is localized in 
one state.
There is no correlation between the
electronic and vibrational 
subsystems at the initial time.
The evolution predicted by Eq.~(\ref{EQ_solution})
is compared with the exact solution~\cite{Jaynes-Cummings63}.
The exact and approximate solutions coincide with good precision until 
$2\pi g t < 1$ holds~\cite{Scully_BOOK},
corresponding to
$t < 6.366$ and $t < 0.6366$ in Figs. 2(a) and 2(b), respectively.
The approximate solution is analytic, in a simple closed form, Eq.~\ref{EQ_solution} 
compared to the exact solution
involving an infinite series summation~\cite{Jaynes-Cummings63}.
Note that
within the time interval determined by $2\pi g t < 1$, the dynamics of inversion has relaxational character 
and usually fits the expression $\frac12 \cos \left( 2\sqrt{\gamma_0}gt\right)\exp{\left(-g^2t^2\right)}$~\cite{Scully_BOOK}.

The sum of the cosine functions in Eq.~(\ref{EQ_solution})
forms beats. 
The fast oscillation of the population inversion
in Fig.~2 
gives excitation transfer  
between the electronic states.
The envelope of the beats describes relaxation in 
the oscillation of the electronic population transfer
due to the vibration induced dephasing. The 
inversion $t_+^{-1}$ and relaxation $t_-^{-1}$ rates 
\bea
t_+^{-1}&=&\frac{\omega_1+\omega_2}{2}, \n \\
t_-^{-1}&=&\frac{\omega_1-\omega_2}{2}.
\label{EQ_inv_rel}
\eea
depend on the number of phonons.
The sum 
and difference
of the frequencies, 
determine the beats of the electronic population 
in Eq.~(\ref{EQ_solution}).
The analytic dependence of 
the rates of inversion (faster component) and relaxation (slower component)
on the vibrational amplitude $\langle \gamma_0\rangle$ and 
electron-vibrational detuning $\delta=\omega-\Omega$ is illustrated in
 Fig.~3. 


The dependence of $t_\pm^{-1}$ on $\delta$ and $\langle \gamma_0 \rangle$ 
becomes clear from the following series expansion:
For $\delta \ll 1$
and 
 $\langle \gamma_0 \rangle \gg 1$ 
the frequencies $t_\pm$
of the population inversion oscillation and relaxation
can be expanded
in powers of $\delta$
and  ${1}/\sqrt{\langle \gamma_0 \rangle}$ 
\bea
t^{-1}_+
        &\simeq&
                 g\sqrt{2\langle \gamma_0 \rangle}
                -\frac{g}{2\sqrt{2\langle \gamma_0 \rangle}}
                +\frac{\delta^2}
                      {2g\sqrt{2\langle \gamma_0 \rangle}}
                +\dots , \n \\
t^{-1}_-
        &\simeq&
                \frac{g}{\sqrt{2}}
                \left(1 + \frac{1}{8\langle \gamma_0 \rangle}\right)
               -\frac{1}{4\sqrt{2}}
                \frac{\delta^2}{g\langle \gamma_0 \rangle}+\dots .
\label{EQ_series_delta}
\eea
The inversion  rate $t^{-1}_+$ has minimum at $\delta=0$,
the relaxation rate $t^{-1}_-$ has maximum at $\delta=0$.
A larger detuning leads to faster oscillations of population and a slower relaxation. 

The dependence of $t_\pm$ on the amount of vibrational energy
is considered in the
$\delta=0$ limit,
which has been explored in Fig.~2. 
Expansion of the sum and difference frequencies up to the third order
in $1/\sqrt{\langle \gamma_0 \rangle}$
with $\delta=0$
takes the form:
\bea
t^{-1}_+
        &\simeq&
                g\sqrt{2\langle \gamma_0 \rangle} 
               -\frac{g}{2\sqrt{2\langle \gamma_0 \rangle}}
               +... , \n \\
t^{-1}_-
        &\simeq&
                \frac{g}{\sqrt{2}}
                \left(1 + \frac{1}{8\langle \gamma_0 \rangle}\right)
               +... .
\label{EQ_series_gamma}
\eea
The frequency of population inversion grows
with system energy 
$\langle \gamma_0 \rangle=\bar \alpha^* \bar \alpha$,
already in the zeroth order of expansion.
The relaxation frequency  decreases as inverse of system
energy, Fig. 3.
Considering the vibrational mode as a quasi heat bath 
with respect to the two-level system,
it is intuitively expected that 
hoter bath
yields quicker relaxation.
The 
decrease of the relaxation rate with increasing phonon energy
occurs since the
oscillator is not in a thermal state, but 
approaches at large $\langle \gamma_0  \rangle$
the classical regime.
A classical oscillator 
coupled to a two state electronic system
yields 
oscillation in the electronic population without relaxation.
For small $\langle \gamma_0 \rangle \le 1$ the
oscillation and relaxation frequences coincide. For large
$\langle \gamma_0 \rangle > 1$, as indicated by expansion (\ref{EQ_series_gamma})
the oscillation frequency grows while the relaxation frequency  decreases.
The frequencies
display branching with energy growth, Fig.~3. 
 
The approximate solution Eq.~(\ref{EQ_solution}) not only gives the
first few oscillations of the electronic
population corresponding to several forward and
backward population transfer reactions, but also well reproduces
the overall dephasing envelope 
that is associated in the thermodynamic limit of many
vibrational modes with relaxation to equilibrium. 
It is quite remarkable that this quasi-equilibration
is observed transiently already with a single
vibrational mode!

%
It is instructive to look further into the dynamics of the population inversion $\left<S_\z \right>$.
The dynamics of this observable is conjugate to the dynamics of the 
expectation value of the correlation operator $\left< \beta \right>$, defined in 
Eqs.~(\ref{EQ_operators}).
As shown by the exact solution
in Fig.~4, 
after a few oscillations the population inversion
stalls at zero indicating that the electronic population is 
equally delocalized between the two states. 
This situation is referred to as 
``collapse''~\cite{cumi65}. 
After a while, the oscillations reappear, 
Fig.~4, 
producing a
``revival'' ~\cite{naro80}. 
It is important to understand in more detail this type of  dynamics,
in particular,
the fate of the energy and phase dynamics
during the ``silent period'', after the collapse and before the revival.  
The electronic contribution to the total energy is stored in an unusual form during 
the silent period. The electronic
population is distributed equally
between the two states. This situation is equivalent 
to the high temperature limit of the populations
of a two-level system coupled to a thermostat.
Here the nontrivial dynamics appear due to the entanglement of 
the electronic and vibrational degrees of freedom.

It is intriguing what happens to the phase of the 
oscillation during the silent period. 
The phase is preserved in a hidden manner to
reappear later when the silent period is over, 
Fig.~4. 
It may be expected that a simple operator or a combination
of several operators exist, whose expectation value
maintains the oscillation, while the oscillation
of the population inversion stalls.
%
The dispersion 
\bea
\sigma_A
        &=&
           \left< \left( 
                         A - \left<A\right> 
                  \right)^2 
            \right>. 
\label{EQ_disper}
\eea
of the vibrational coordinate
 $(A=q)$
is the desired expectation value. The vibrational
dispersion starts at the minimal value at time zero,
grows with time and maintains the oscillation
during the silent period, Fig.~5. 

The origin of the oscillation in the dispersion of the vibrational coordinate is illustrated by the evolution
of the vibrational wavepacket in Fig.~6 
that shows wavepacket snapshots at
several times. Initially, the wavepacket is
Gaussian. As time evolves, the electron-vibrational
coupling splits the vibrational wavepacket 
into two branches
 correlated with the  two electronic states.
The dispersion of the vibrational coordinate 
represents the width of the overall wavepacket.
The width oscillates as the two wavepacket
branches evolve nearly independently
as determined by their own electronic states.
The width of the overall wavepacket is determined
by the separation between the branches.
The width is maximal when the two wavepackets 
are far apart and minimal when 
the two wavepackets penetrate through each other. 
After the silent period when
the population inversion resumes its oscillation,
the wavepacket width oscillation decreases in amplitude, Fig.~6. 
 

\section{Analysis of uncertainty} \label{Analysis_of_uncertainty} 

Dephasing of the electronic subsystem
due to coupling to vibrations is ubiquitous in photochemistry.
The silent period 
                    when the dynamics 
                    of the electronic population
in the Jaynes-Cummings model
                   temporarily stops 
due to the vibrationally induced dephasing
                   is investigated further.
Consider
the uncertainties of the observables 
describing the vibrational and electronic modes.
The vibrational wavepacket 
with two peaks that exist during the silent period, 
Fig.~6, 
has a large uncertainty.
As shown before, 
the phase of the electronic subsystem 
is preserved in a hidden form 
by the uncertainty of the vibrational coordinate.


The uncertainty relationship 
for dispersions $\sigma_A$, $\sigma_B$, 
Eq.~(\ref{EQ_disper}), 
of Hermitian operators $A$ and $B$ 
is commonly written in a product form as~\cite{Berman_Gena_BOOK}
\bea
\sigma_A 
\cdot 
\sigma_B 
         &\ge& \left|
                \frac14 \left< [A,B]\right>^2
               \right|.
\label{EQ_disper_product}
\eea
An alternative relationship between
the dispersions in a sum form can be obtained
based on the positivity of the norm of operator
$F= A-\langle A \rangle + i \left( B-\langle B\rangle\right)$, 
$\langle F^* F \rangle \ge 0$
$\to$ 
$\langle
   \(   A-\langle A \rangle                       \)^2
+  \(   B-\langle B \rangle                       \)^2
+ i\[   A-\langle A \rangle, B-\langle B \rangle  \]
\rangle \ge 0$, and
\bea
  \sigma_A 
+ \sigma_B 
+ i\left< [A,B] \right> 
                       &\ge& 0.
\label{EQ_disper_sum}
\eea
The sum 
        $ \sigma_A + \sigma_B + i\left< [A,B] \right>$ 
     gives a good second-order combination of operators, 
     whose expectation value 
     can be included into the chain  
     of Heisenberg EOMs.
In application to the vibrational subsystem, 
the sum uncertainty relationship~(\ref{EQ_disper_sum}) means that 
the sum of dispersions for
the vibrational position and momentum
\bea
A &=&    a^+ + a    = q \sqrt{2m\omega/\hbar} \n \\
B &=& i\(a^+ - a \) = p \sqrt{2/(\hbar m \omega)}
\label{EQ_bosonic_uncertainty}
\eea
must be larger than zero 
\bea
 \left<a^+a\right>
-
 \left<a^+\right>\left<a\right> 
 &\ge& 0
\label{EQ_disper_bosonic}
\eea

Consider the dispersion of coordinate
in more detail. It can be written in terms
of the raising and lowering operators
in the form 
\bea
\label{EQ_vibrational_dispersion}
\sigma_q
          &=&
             \left<   \(q-\left<q\right>\)^2 \right>  \\
          &=&
             \frac{\hbar}{2m\omega}
             \left< \[ a^++a - \left<a^++a \right> \]^2\right> \n\\
          &=&  
            \frac{\hbar}{2m\omega}
         \left\{ \right.
          \underbrace{
                \left<aa^+\right>-\left<a\right>\left<a^+\right>   
              + \left<a^+a\right>-\left<a^+\right>\left<a\right>   
                     }_{{\rm cross~term}}
          \underbrace{
              +\left<\(a^+\)^2\right>-\left<a^+\right>^2 \n 
              +\left<a^2\right>-\left<a\right>^2 
                     }_{\rm ladder~term}
         \left. \right\} . \n
\eea
The coordinate dispersion includes two kinds
of terms, the ladder 
dispersion terms 
involving operator squares that increase or
decrease the number of quanta by two,
and the cross terms including 
products of conjugate ladder operators that
preserve the number of quanta.  
The mean value of the ladder term 
oscillates quickly and 
gives
fringes 
in Fig.~5. 
The cross term 
maintains the rotating wave approximation
preserving the number of quanta, and
evolves slowly.
Similar decomposition applies
to the dispersion of momentum
\bea
\label{EQ_disp_momentum}
\sigma_p
        &=& 
             \frac{m\omega\hbar}{2}
              \langle
                \left[
                     a^+-a -\left<a^+-a\right> 
                \right]^2  
			  \rangle \\
        &=&  \frac{m\omega\hbar}{2}
              \left\{
            \underbrace{
               \left<aa^+\right>-\left<a\right>\left<a^+\right> 
              +\left<a^+a\right>-\left<a^+\right>\left<a\right> 
                       }_{\rm cross~term}
           \underbrace{
             -\[
                \left<\(a^+\)^2\right>-\left<a^+\right>^2 
             +  \left<a^2\right>-\left<a\right>^2  
              \] 
	                  }_{\rm ladder~term}
               \right\} . \n
\eea
The difference from the position operator
dispersion is only in the negative sign by the ladder term.
The quickly oscillating ladder terms cancel
out in the sum of the coordinate and momentum dispersions, 
leaving only the cross terms 
\bea
\frac14
\left(
 \sigma_A
+\sigma_B
\right)
             &=&
                   \frac{m\omega}{2\hbar}\sigma_q
                 + \frac{1}{2m\omega\hbar}\sigma_p 
              =    \left\{
                           \underbrace{
                            \langle a^+a \rangle +1/2
                                      }_{\rm total~energy}                             
                         - 
                           \underbrace{
                            \langle a^+\rangle \langle a \rangle 
                                      }_{\rm semiclassical}
                    \right\}
             = E^{\rm vib}_{\rm quant}.
\label{EQ_sum_disp}
\eea
The slowly oscillating
sum of dispersions~(\ref{EQ_sum_disp})
is nothing but a purely quantum contribution 
to the vibrational energy defined as the 
difference between
the total vibrational energy
\bea
E_{\rm total}^{\rm vib}
             &=&
                \frac{1}{2m}        \langle p^2 \rangle
              + \frac{m\omega^2}{2} \langle q^2 \rangle
              =
                \hbar \omega \langle a^+ a + 1/2 \rangle,
\label{EQ_vib_total}
\eea
and the semiclassical contribution to the total energy
\bea
E_{\rm semi-cl}^{\rm vib}
               &=&
                \frac{1}{2m}        \langle p \rangle^2
              + \frac{m\omega^2}{2} \langle q \rangle^2
              =
                \hbar \omega \langle a^+  \rangle \langle a \rangle,                  
\label{EQ_vib_class}
\eea 
formed of the first order 
classical-like averages 
$\langle q \rangle$ and $\langle p \rangle$.
The purely quantum contribution to the vibrational energy 
provides the envelope of the coordinate dispersion 
in Fig.~5. 
Substitution of the
 commutation relation
$\left[p,q\right]=i\hbar$
$\to$
$       \[ A, B\]=
                 \[  q \sqrt{2   m \omega /\hbar}  ,  
                     p \sqrt{2 / m \omega  \hbar} \]
             =   2i$
and the sum of dispersions~(\ref{EQ_sum_disp})
into the general formula~(\ref{EQ_disper_sum})
 gives the
uncertainty relationship~(\ref{EQ_disper_bosonic})
for the vibrational dispersions.

Next, consider the sum uncertainty relationship,
Eq.~(\ref{EQ_disper_sum}),
for the electronic, generally fermionic subsystem
with
\bea
A &=& \frac12 \(S_+ + S_- \)  = S_\x , \n \\
B &=& \frac i2 \(S_+ - S_- \) = S_\y .
\label{EQ_disper_fermionic}
\eea
The inequality~(\ref{EQ_disper_sum}) gives 
\bea
 \left< S_+ S_- \right>
- \left<S_+\right>\left<S_-\right>  
\ge 
     0.
\label{EQ_fermionic_uncertainty}
\eea
Similarly to the vibrational subsystem,
the dispersions of the electronic 
$S_\x$ and $S_\y$ operators 
can be decomposed into the
cross and ladder terms. 
The dispersion
$\sigma_{S_\x}$ reads
\bea
\label{EQ_electronic_dispersion}
\sigma_{S_\x}
         &=&
             \frac14
             \left< \[S_++S_- - \left<S_++S_- \right> \]^2\right> \\
         &=&
            \frac14
            \left\{ \right.
          \underbrace{
            \left<S_+S_-\right>-\left<S_+\right>\left<S_-\right> 
          + \left<S_-S_+\right>-\left<S_-\right>\left<S_+\right> 
                     }_{\rm cross~term}
          \underbrace{
          + \left<S_+^2\right>-\left<S_+\right>^2               
          + \left<S_-^2\right>-\left<S_-\right>^2
                     }_{\rm ladder~term}
            \left. \right\}. \n
\eea
Since the second order averages in the last equation add to a constant, 
$\left<S_+S_-+S_-S_+\right>=1$ by
completeness, 
and 
$\left<S_+\right>^2=\left<S_-\right>^2=0$
due to finite dimensionality 
of the electronic basis, 
the electronic dispersion involves
only the first order averages 
\bea
\sigma_{S_\x}
       &=&
           1/4
               \left(
               \underbrace{
                  1-2\left< S_+ \right> \left< S_- \right> 
                          }_{\rm cross}
               \underbrace{
                -\left<S_+\right>^2
                -\left<S_-\right>^2
                          }_{\rm ladder}
               \right).
\label{EQ_short_disp}
\eea
The ladder terms oscillate quickly and are
responsible for the 
fringes in Fig.~7. 
Similarly, the dispersion of $S_\y$ 
can be decomposed into the cross and ladder terms
\bea
\sigma_{S_\y}
       &=&
           1/4
               \left(
               \underbrace{
                  1-2\left< S_+ \right> \left< S_- \right> 
                          }_{\rm cross}
               \underbrace{
                +\left<S_+\right>^2
                +\left<S_-\right>^2
                          }_{\rm ladder}
               \right).
\label{EQ_short_disp_y}
\eea
The quickly oscillating ladder terms cancel out
in the sum of the two dispersions
\bea
 \sigma_{S_\x}
+\sigma_{S_\y}
               &=&
                  \frac12
                  -\left< S_+ \right> \left< S_- \right> .
\label{EQ_short_el_sum}
\eea
The  sum of the electronic dispersions up to a constant and a sign
is nothing but
the quasiclassical contribution
\bea
E_{\rm quasicl}^{\rm el}
    &=&
        \hbar \Omega \langle S_+ \rangle \langle S_- \rangle
\label{EQ_el_classic}
\eea
to the total electronic energy
\bea
E_{\rm total}^{\rm el}
   &=&
              \hbar \Omega \langle S_+  S_- \rangle .
\label{EQ_el_total}
\eea
Note that the corresponding sum of vibrational dispersions gives the quantum contribution to vibrational energy, Eq.~(\ref{EQ_sum_disp}).
The difference between $E_{\rm quasicl}^{\rm el}$ and $E_{\rm total}^{\rm el}$ 
then is the purely quantum contribution 
to the electronic energy
\bea
E_{\rm quantum}^{\rm el}
       &=&
            E_{\rm total}^{\rm el}
          -
            E_{\rm quasicl}^{\rm el}
        = 
              \hbar \Omega \left( 
                                \langle S_+  S_- \rangle 
                               -\langle S_+ \rangle \langle S_- \rangle
                           \right).
\label{EQ_el_quantum}
\eea
Substitution 
of the commutator
$\left[S_\x,S_\y\right]=-iS_\z$ 
and sum of electronic dispersions 
Eq.~(\ref{EQ_short_el_sum}) 
into the general formula~(\ref{EQ_disper_sum})
gives the
relation~(\ref{EQ_fermionic_uncertainty})
between the electronic dispersions
in an equivalent form
\bea
\label{EQ_el_for_figure}
\sigma_{S_\x}+\sigma_{S_\y}+\langle S_\z \rangle \ge 0 ,
\eea
with
$S_\z=S_+S_--1/2$
and
Eq.~(\ref{EQ_short_el_sum}). 
Eqs.~(\ref{EQ_fermionic_uncertainty})
and (\ref{EQ_el_for_figure})
imply
that the purely
quantum
part of the electronic
energy
is always greater than zero, 
$E_{\rm quant}^{\rm el} \ge 0$.
Over time the
electronic  energy  flows between its quasiclassical and
quantum parts, as shown in Fig.~7. 
The quasiclassical contribution to the electronic energy provides
an envelope for the electronic dispersion.


The electronic and vibrational dispersions show interesting similarities that are summarized in Table~I.
The fringes in the electronic dispersion, Fig.~7, 
appear at the same phase and frequency as the fringes in the vibrational dispersion, 
Fig.~5. 
The difference is in the sign of the deviation of the electronic and vibrational dispersions 
from their bounds, cf. the signs in inequalities~(\ref{EQ_disper_bosonic}) and~(\ref{EQ_fermionic_uncertainty}).
The fringes in the vibrational dispersion are directed upward, 
while the fringes in the electronic dispersion are directed downward.
The electronic dispersion is bounded from above in contrast to the vibrational dispersion, which is bounded from below.

 Figure 8 
displays the sums of dispersions for the vibrational and 
subsystems in comparison to
the real valued electron-vibrational correlation 
\bea
i
\langle
\beta_r
\rangle
        &=&
            \left\{
                    \langle  \beta \rangle
                  - \langle  a   \rangle \langle S_+ \rangle
                  - \langle  a^+ \rangle \langle S_+ \rangle
            \right\},
\label{EQ_correlator}
\eea
directly related to the auxiliary operator $\hat \beta$, Eq.~(\ref{EQ_operators}).
The correlation oscillates with the same low frequency as the electronic and vibrational dispersions.
Rigorously,
the combination of operators Eq.~(\ref{EQ_correlator}) can naturally
appear as a sum of dispersions $\sigma_A + \sigma_B$ for
cross-operators, composed of both fermionic and bosonic parts 
$A=S_++ a$, $B=i\left( S_+ - a\right)$ 
and their conjugates.
For the initial condition with
no correlation between the fermionic and bosonic
modes, and zero horizontal spin projections  $\langle S_+\rangle\left.\right|_{t=0}=0$, the
expectation value of Eq.~(\ref{EQ_correlator}) starts at zero and
oscillates around zero.  
Up to zero point value and scaling constants,
the dispersions of electronic subsystem 
and the correlation Eq.~(\ref{EQ_correlator})
behave as sine and cosine functions.

When the population inversion stalls, Fig.~4, 
the electron-phonon system is in a 
 state with large uncertainty. 
For both electronic and vibrational subsystems,
states with larger uncertainties
provide
larger purely quantum energy contributions.
As shown in Fig.~5, 
 the dispersion 
associated with the vibrational subsystem  increases, 
in accord with the uncertainty relationship~(\ref{EQ_disper_bosonic}). 
In contrast 
to the vibrational dispersion, 
the dispersion associated with the electronic subsystem 
decreases
as seen in Fig.~7, 
in accordance with 
the uncertainty relationship~(\ref{EQ_fermionic_uncertainty}).
The electronic subsystem 
approaches the state of minimal uncertainty 
when inversion oscillation stops.

The unusual behavior of the dispersion associated with the electronic subsystem 
can be understood with the concept of pure and mixed states.
An isolated electronic system is
characterized by either a  normalized wave function 
$\ket{\psi}
          =
               c_0\ket{0}+c_1\ket{1}$,
or a density matrix,
 or quantum mechanically averaged values of electronic operators.
In an entangled electron-phonon system 
the electronic part
is a statistical average 
over different realizations of the vibrational subsystem.
The statistical mixture of states cannot be represented through a wave function,
and is characterized  by a reduced density matrix
or both statistically and quantum mechanically averaged values of electronic operators.

The 
density matrix of an isolated electronic system
is determined by the
 products of the wave function coefficients
$\rho_{\rm el}
               =
                         \ket{\psi}\bra{\psi}
               =
                 c^*_1c_1\ket{1}\bra{1}
             +   c^*_1c_0\ket{1}\bra{0}
             +   c^*_0c_1\ket{0}\bra{1}
             +   c^*_0c_0\ket{0}\bra{0}
               $.
The reduced density matrix of the electronic subsystem of
 the electron-phonon system
can be specified by
expectation
values of electronic operators
averaged over the state of the whole electron-phonon system
$\rho_{\rm el}
               =
                \langle  S_+S_-\rangle   \ket{1}\bra{1}
              +  \langle    S_+ \rangle  \ket{1}\bra{0}
              +  \langle    S_- \rangle  \ket{0}\bra{1}
              +  \langle  S_-S_+\rangle  \ket{0}\bra{0}
               $.
The trace of the squared density matrix   of an isolated electronic system
is a complete square and always equals unity
${\rm Tr} 
     \rho
     _{\rm el}^2
                \equiv
                  \left( c_1^*c_1 + c_0^*c_0\right)^2 = 1$,
due to normalization of the wave function.
The trace of the reduced electronic density matrix squared, often referred to as fidelity, 
takes the form
\bea
{\rm Tr}
\rho^2
_{\rm el}
          &=&
              1             
               +2\langle S_+\rangle \langle S_-\rangle 
              - \langle S_+S_-\rangle\langle S_-S_+\rangle
         \le 1,
\label{EQ_fidelity}
\eea     
and is less or equal one.
The last term is nonnegative, because it is a product of nonnegative populations
$\langle S_\pm S_\mp \rangle \ge 0$.
The second term is zero if 
the averaging over the phonon state gives
 no correlation between
$\ket{0}$ and $\ket{1}$.
${\rm Tr}\rho^2_{\rm el}=1$ means   that the electronic subsystem exists in a pure state.
${\rm Tr}\rho^2_{\rm el}<1$ implies that the electronic subsystem     is in a mixed state.

The distinction between pure and mixed states is best illustrated with
the Bloch vector
\bea
\vec{R}\equiv(\left<S_\x\right>,\left<S_\y\right>,\left<S_\z\right>). 
\label{EQ_bloch}
\eea
The square of the Bloch vector $R^2$ is related to ${\rm Tr}\rho^2_{\rm el}$
by a linear transformation given in Table II.
The Bloch vector~(\ref{EQ_bloch})
connects the origin with the point in a three-dimensional space
specified by the expectation values of the projections of the spin-1/2 operator onto x, y, z axes.
\bea
R^2
   &=&
       \left<S_\x\right>^2+\left<S_\y\right>^2+\left<S_\z\right>^2 \le 1/4.
\label{EQ_Bloch_sphere_radius}
\eea
The square of the Bloch vector can be decomposed
$R^2=\langle S_\z \rangle^2+R_{\x\y}^2$
into the
 vertical $\langle S_\z\rangle$
and horizontal components
\bea
R
_{\x\y}
^2
       = \left<S_\x\right>^2+\left<S_\y\right>^2
        =  \left<S_-\right>\left<S_+\right>
.
\label{EQ_horisontal}
\eea
The x, y, and z components of the Bloch vector of an
isolated electronic system are
$\left(c^*_1c_0+c_0^*c_1\right)/2$,
$\left(c^*_1c_0-c_0^*c_1\right)/2i$,
$\left(c^*_1c_1-c_0^*c_0\right)/2$
yielding
$R^2
     =
      \frac14
      \left[
            \left( c^*_1c_1\right)^2 
            +\left( c^*_0c_0\right)^2 
            +2 c^*_1c_1c^*_0c_0
     \right]=1/4$.
The squared length of the Bloch vector
of the electronic subsystem of a coupled electron-phonon system is
\bea
R^2
    &=&
        1/4
      + \langle S_+\rangle  \langle S_-\rangle
      - \langle S_+S_-\rangle\langle S_-S_+\rangle \le 1/4.
\label{EQ_Bloch_ladder}
\eea
Here, the last term is always nonnegative, and the second term is zero 
if the correlation between 0 and 1 is dephased by averaging over the vibrational states. 
$R^2$ decreases to zero  
when the electronic subsystem is equally mixed between the two states, $\left<S_+S_-\right>=\left<S_-S_+\right>=1/2$, 
 the last term compensating the 1/4 constant. 
The relationships between the square of the Bloch vector 
and the trace of the squared density matrix are summarized in Table II.  
As shown in Fig.~8(c) 
for electronic subsystems with equal populations of both electronic states $\langle S_\z \rangle=0$,  
$\langle S_+S_- \rangle = \langle S_- S_+ \rangle=1/2$,  
the change of the squared Bloch vector from 0 to 1/4 corresponds to the change of fidelity from 1/2 to 1. 
Fidelity expressed in terms of the electronic dispersions 
shows that mixed electronic states correspond to 
 larger values of the electronic dispersions. 
Pure electronic states yield 
minimal uncertainties of electronic operators,  
minimal values for the electronic dispersions 
and the maximal length of the Bloch vector R=1/2. 
Alternatively, mixed electronic states yield 
larger electronic operator uncertainties, 
larger electronic dispersions 
and shorter Bloch vectors.

The end of the Bloch
vector glides 
from the
south pole 
of the Bloch sphere 
to the north pole 
then back to the south pole  and so on.  
During this revolutions, shown in the lower panel of Fig.~9, 
the length of the Bloch vector decreases, 
as the 
electronic subsystem becomes
 a statistical mixture. 
%
%
The mixing is maximized at $\omega t \simeq 1.5-2.0$.
For times $\omega t > 2$
the Bloch vector 
becomes larger
and  the electronic subsystem returns closer to a pure state.
The length 
of the Bloch vector never recovers its initial value during the studied time period.
In the middle of the silent period 
the Bloch vector is almost at maximal length $ R^2 \simeq 1/2 $.
Its
z-axis projection 
is zero
$\langle S_\z \rangle =0$
in accordance with Fig.~4. 
Thus, during the silent period the Bloch vector revolves equatorially (nutates) around z-axis.

The analysis of fidelity shows that the electronic subsystem 
successively passes through the following sequence of states. 
Initially in a pure state, the electronic subsystem quickly becomes 
a mixed state with a small fidelity. 
This occurs within one-eighth of the revival period, 
defined by 
$\Omega_{\rm revival} \simeq g/\sqrt{\gamma_0}$
\cite{naro80}.  
Then, the fidelity rises and the electronic subsystem again closely approaches a pure state. 
The fidelity is maximized at exactly one half of the revival period, $t=\pi/\Omega_{\rm revival}$.

During the silent period 
when the electronic population stalls, 
the oscillation shifts to the vibrational and electronic dispersions and 
$\langle S_\x \rangle$, $\langle S_\y \rangle$ expectation values that preserve the phase of the oscillation 
and are maximized exactly in the middle of the silent period. 
Generally, the electronic and vibrational dispersions and electronic coherences $\langle S_\x \rangle$, $\langle S_\y \rangle$ 
are maximized at half integer revival times 
$t_{\rm revival}=2\pi/\Omega_{\rm revival}=2\pi \sqrt{\gamma_0}/g$, 
 while the electronic inversion $\langle S_\z \rangle$ 
and vibrational classical-like $\langle p \rangle$ and $\langle q \rangle$ are maximized at integer revival times.

\section{Time scales} \label{Time_scales}   

The coupled electron-phonon dynamics in the molecular dimer forms a hierarchy of time scales. 
The electronic and vibrational coherences $\langle S_\x \rangle$ and $\langle q \rangle$, 
as well as electronic population $\langle S_\z \rangle$ 
 and energy $ \langle a^+a \rangle$ oscillate with different frequencies that vary by order of magnitude or more.
 Each frequency component can be associated with a distinct physical process.  
The following is the hierarchy of the time scales in increasing order. 
The quickest time scale $t_I$
 corresponds to the oscillations of the expectation values of the vibrational coordinate 
$\langle q \rangle$
 and electronic coherence $\langle S_\x \rangle$, $\langle S_\y \rangle$. 
The second time scale $t_{II}$
characterizes
 the inversion of the electronic population $\langle S_\z \rangle$. 
The population inversion time depends on the initial displacement of the vibrational wavepacket. 
Larger displacements yield quicker population transfer. 
Relaxation and dephasing of the population inversion occurs on a slower time scale $t_{III}$. 
Over this time interval the electronic state approaches quasi-equilibrium 
with equal partitioning of the population between the two states. 
The slowest component $t_{IV}$ is associated with the silent period 
and determines the revival of the oscillation in the electronic state populations. 
The $t_{IV}$ time also determines 
the time scale 
of the relaxation 
of the expectation value
of the vibrational coordinate $\langle q \rangle$ to its equilibrium position. 
The electron-vibrational system enters and leaves the state with the maximal uncertainty on the $t_{IV}$ time scale.

The time scales of the faster processes $I$ through $III$
 can be estimated from the approximate analytic expressions 
(\ref{EQ_solution})-(\ref{EQ_frequencies})
 derived in this work. 
The slowest time scale $IV$ is estimated by Eberly {\it et al.}~\cite{naro80}
 and is reproduced in our numerical simulations.  
The slowest time scale $IV$ describing the silent period and revivals
 of the oscillation in the electronic populations 
does not appear as a separate time scale in the analytic expressions 
(\ref{EQ_solution})-(\ref{EQ_frequencies})
 and coincides with the time scale $III$ of the relaxation and dephasing of the oscillation.  
The revivals in the electron-phonon system can take longer 
than the dephasing time and produce a silent period. 
According to Eberly {\it et al.}~\cite{naro80}, 
$t_{IV}$ is determined by the difference in the Rabi flopping frequencies
 of the two neighboring and most populated levels of the oscillator. 
The following are the rates associated with the four time scales
\bea
t_{\rm    I}^{-1} &\sim&         \omega_\vi, \n \\
t_{\rm   II}^{-1} &\sim& g \sqrt{2\left<\gamma_0\right>}, \n \\
t_{\rm  III}^{-1} &\sim& g /\sqrt{2} ,                 \n \\
t_{\rm   IV}^{-1} &\sim& \frac{g}{\sqrt{\left<\gamma_0\right>}}.
\label{EQ_time_scales}
\eea
Even for this simple two dimensional system isolated from environment, 
the time taking the system to return to its initial state, 
the Poincare recurrence time can be very long. 
The quasi-irreversibility feature appears 
due to discreteness of the oscillator states
and determines the character of the long-time dynamics. 
 After each revival period the state of the system is farther away from the initial state. 
For most real systems with large number of degrees of freedom even the first revival is barely accessible. 
According to Figures 2 
and 5, 
after a long period of time 
the expectation values of the population inversion and vibrational coordinate 
lose the revivals and approach quasi-equilibrium. 
Figure~7 
indicates that some electronic coherence may persist for a long time. 
Based on the above analysis it is reasonable to expect that 
after a sufficiently long time interval 
the system will be found in a state of large uncertainty 
with the first order dynamical averages performing small 
amplitude chaotic motions about the equilibrium values.

 \section{Discussion}  \label{Discussion}
Few simple quantum models  allow for analytic treatment: 
two interacting harmonic oscillators 
as well as
two interacting two-level systems are exactly solvable. 
A harmonic oscillator interacting with a two-level system 
cannot be solved exactly by use of a finite number of variables.
Known are the infinite series solutions of 
the eigenstate method and finite approximations by the QHD method, presented here.
The  infinite set of wave function coefficients in the equations of motion 
in the eigenstate solution is equivalent to the infinite chain of QHD equations. 

The dynamics of 
a massive particle in a harmonic potential, 
undergoes the same evolution as a quantized mode of a massless electromagnetic wave.
The relaxation of the population inversion is irreversible in general,
so that the dynamics can be referred to as reversible decoherence.
An interesting issue is observed in Fig. 6(a) at $\omega t/(2 \pi)=9.5$. 
The dispersion of coordinate closely approaches its initial value.
The states of the electronic and vibrational subsystems become correlated with time. 
This nontrivial behavior of the averages 
is associated with formation of  entanglement.

Applications and extensions of this work cover three directions: 
theoretical, chemical, and spectroscopic.
From the theoretical point of view the results can be 
improved by using  higher order closure relations.
The necessity to take into account zero-point energy effects and get a better description 
of the silent period suggests an  extension of the
calculation scheme by including equations of motion for new variables,
in particular, the dispersions of products of nuclear and electronic operators. 
Variables which follow state-specific positions, 
electronic phase, and vibrational phase acquired by the dimer with time
can be included into the calculation scheme
for application to echo experiments. 
For instance, the state-specific positions are $\langle q \left( S_\z \pm 1/2 \right)\rangle$.
The relative electronic phase 
can be defined with
$\langle S_\x \rangle$, $\langle S_\y \rangle$.
The vibrational phase is specified by a variable, analogous to the $\gamma$ term  in the Heller work on Gaussian wavepackets~\cite{Heller70}. 
 Some of these new variables and equations of motion appear
naturally while describing the coupled electron-phonon dynamics 
with the Fokker-Planck equation of motion 
for the normally ordered 
multidimensional generating function, 
comprising both electronic and nuclear degrees of freedom~\cite{Carmichael_BOOK}.

An important chemical application 
of the presented method 
of calculating the coupled 
electron-phonon dynamics 
is the description 
of energy and charge transfer 
in large molecular aggregates
 playing important biological and technological roles, 
 e.g. in the light harvesting complex 
and molecular solar cells. 
The dynamics of 
electronic  excitations 
is coupled in large systems to many inter- and intramolecular 
nuclear degrees of freedom 
as well as to 
vibrations of environment,
whose dynamics are extremely difficult to follow on  the quantum level.
The QHD method used here follows the quantum properties using few expectation values 
to characterize each degree of freedom, and describes the exciton and electron  transfer dynamics in a simple way. 
The QHD 
method is directly applicable to the 
quantitative description of transfer process 
in large chemical systems.

The method of this paper is also able to describe
3 pulse-, 4 pulse-, and correlated photon echo spectroscopic experiments with molecular systems
by addition of  new variables.
In femtosecond echo experiments with ensembles of disordered 
dimers~\cite{Grahm_Fleming_AND_Xantippe,Freming3PEPS}, 
the evolution of phase 
for each dimer in the ensemble determines the re-phasing of the ensemble 
net transition dipole and, consequently, the photon echo  signal.
In particular, for the three-pulse stimulated photon echo 
technique, 
the phase acquired during the population period between pulses II and  III  
determines the time delay for appearance of the stimulated echo.
The echo signal determined by the relative phase
 between the ground and one of the excited electronic states, 
 carries information about the population transfer between the 
optically active excited states. 
It is expected that 
oscillations in the third-order echo peak shift, modulated by the damped evolution of the excited states population $\langle S_\z \rangle$, 
will be reproduced by the current method.
In a  four pulse femtosecond experiment, 
the fourth pulse 
serves for the heterodyning: The excited state wavepacket is projected onto the ground state wavepacket. 
The created interference population  in the excited state  affects the net fluorescence from this  state. 
The time of the signal is determined by the acquired electronic phase 
and  position of the exciton transfer wavepacket described by the $\langle q \left( S_\z \pm 1/2 \right) \rangle$ variable.

A novel correlated spectroscopic technique,
based on a simultaneous irradiation of a sample by pulses of 
optical (electronic)
and IR (vibrational)
 frequencies
can yield new information about the dynamics of 
the electron-vibrational
correlation
c.f. $\langle \alpha \rangle$ and  $\langle \beta \rangle$, calculated in this paper.
We propose to excite and probe by correlated infrared and optical pulses.
The scenario is analogous to the 2D IR spectroscopy used to get the coupling 
between the vibrational modes with close frequencies~\cite{Tokmakoff}.
For the case of symmetric modes, the dynamics of coupling
between optically active low frequency vibrational and high frequency electronic modes
 can be revealed by 2D Raman.
For non-symmetric modes direct and independent excitation of the vibrational degree of freedom with an IR-pulse 
and the electronic degree of freedom with an optical pulse is required. 
The technical difficulty of this experimental method is establishing  the phase relations 
between the IR and optical excitations. 
The experimental data measured by such technique can be modeled by the method developed in this paper.

\section{Conclusions} \label{Conclusions}

Quantum dynamics 
of an electron-nuclear excitation 
in a model dimer has been analyzed in this work
in the limit of small nuclear reorganization. 
The relaxation behavior in the model 
has been obtained by a reduced description that, 
by simplicity, is applicable to large systems. 
The approximate approach quantitatively reproduces
the correct behavior of the model. 
It is also demonstrated that even in the absence 
of the heat bath the system approaches 
a mixed state corresponding to a quasi-thermal equilibrium, 
by destructive interference 
of wave function coefficients.

The influence of the vibrational mode onto the electronic 
population dynamics is analyzed by solving a chain of  coupled 
differential equations for 
population inversion, 
electron-vibrational correlation, etc.
The derived analytic result 
describes oscillation and relaxation 
of electronic population at short times. 
The approximate solution correctly represents
the first several population inversions
and the  decay in the amplitude 
of the oscillation of the electronic population. 
The approximation works better for the quasiclassical 
regime with large vibrational amplitude and weak coupling.

Detuning of the electronic and vibrational frequencies 
slows down the relaxation of the electronic population oscillation.
An increase of the initial vibrational energy qualitatively changes 
the evolution of the electronic population
 from continuous oscillation to beats that reflect relaxation.
The rate of the relaxation 
drops with increasing 
vibrational amplitude.
It is found that
several
dynamical variables oscillate
 with the same rate and relax at the same time. 
Quantum expectation values, such as the  vibrational coordinate dispersion
not included into the primary calculation scheme, 
preserve the dynamics during the silent period,
when
a single vibrational mode acts as a quasi heat bath 
and induces electronic population dephasing over a finite time.
The relaxation 
of the electronic population and a noticeable growth 
of the vibrational 
coordinate dispersion
originate from splitting 
of the vibrational  wave function into two wavepackets that
oscillate independently and correlate with different electronic states.
The original Gaussian wavepacket 
transforms 
 into a non-classical state with large width.
The vibrational energy flows between 
its quasiclassical and purely quantum components.
Relaxation of the oscillation of the overall vibrational energy 
is compensated by growth in the oscillation of the vibrational coordinate dispersion.

The vibrational energy contains more quasiclassical energy at the initial and revival times
 and less quasiclassical energy during the silent time interval between the revivals.
At the time of revivals the vibrational subsystem comes into a quasiclassical state.
By analogy, 
the electronic energy flows
 between its quasiclassical and purely quantum components.
The electronic energy contains less quasiclassical 
contribution at revival times and more quasiclassical contribution during the silent period.
The interaction with the vibrational mode 
drives the electronic subsystem into a mixed state and reduces the length 
of the Bloch vector. In the middle of the silent interval,
the electronic subsystem comes into a coherent state.
This qualitative process is reflected in the time evolution of the
squared length of the Bloch vector, the sum of the 
electronic dispersions, and the fidelity of the electronic subsystem,
which are all linearly dependent.

Time dependence of 
the sums of the electronic and vibrational dispersions, 
and the electron-vibrational correlation oscillate with a 
low frequency corresponding 
to the rate of relaxation and dephasing of the fast 
first order dynamical variables.
The extrema of the sums of the electronic and vibrational dispersions correspond 
to the fastest change in the electron-vibrational correlation and vice versa.
Oscillations of the dispersions of the electronic
and vibrational operators occur with the same
frequencies.
The 
correlation 
between the electronic and vibrational subsystems 
grows, oscillates, relaxes, and reappears 
similarly  to the population inversion. 
Work in progress extends the reported results
toward more accurate calculations 
and descriptions of molecular charge and excitation transfer
observed in ultrafast optical experiments.

\section*{Acknowledgments}
The research was supported by NSF, CAREER Award CHE-0094012. OVP is a
Camille and Henry Dreyfus New Faculty and an Alfred P. Sloan Fellow.
DSK thanks A. Piryatinski for fruitful discussions.



\begin{thebibliography}{10}
\bibitem{Zewail} A. H. Zewail, J. Phys. Chem. A {\bf 104} 5660 (2000); Science {\bf 242} 1645 (1988).
\bibitem{balzani_scandola}          V.~Balzani and F.~Scandola, {\it Supramolecular Photochemistry} (Ellis Horwood,  Chichester, 1991).
\bibitem{Rienk_van_Grondelle-BOOK} H. van Ameroningen, L. Vakunas, 
				and R. van Grondelle, 
				Photosynthetic Excitons (World Scientific, 2000).
\bibitem{monat02} J.~E.~Monat, J.~H.~Rodriguez, J.~K.~McCusker,  J. Phys. Chem. A {\bf 106} 7399 (2002).
\bibitem{stier03} W.~Stier and O.~V.~Prezhdo, J.~Phys. Chem. B {\bf 106} 8047 (2002); 
              Isr. J. Chem. {\bf 42} 213-224 (2003); J. Mol. Struct. (Theochem.) {\bf 630} 33 (2003);
              W.~Stier, W.~R.~Duncan, and O.~V.~Prezhdo, Adv. Mat., {\it in press}.
\bibitem{brook03} C. Brooksby, O.~V.~Prezhdo, and P.~J.~Reid, J. Chem. Phys. {\bf 118} 4563 (2003); {\it ibid} {\bf 119} 9111 (2003).
\bibitem{nobel_prize88} J. Deisenhofer and  H. Michel, Annu. Rev. Biophys. Biophys. Chem. {\bf 20}, 247-266 (1991). 
\bibitem{marin02} R.~Uberna, M.~Khalil, R.~M.~Williams, J.~M.~Papanikolas, and S.~R.~Leone, J. Chem. Phys. {\bf 108} 9259 (1998);
	     		   T.~W.~Marin, B.~J.~Homoelle, K.~G.~Spears, J.~T.~Hupp, and L.~O.~Spreer, J. Phys. Chem. A {\bf 106} 1131 (2002).
\bibitem{Grahm_Fleming_AND_Xantippe}  G.~D.~Scholes, X.~J.~Jordanides, G.~R.~Fleming, J. Phys. Chem. B     
 						 {\bf 105} 1640 (2002).
\bibitem{wasi92} M.~R.~Wasielewski, Chem. Rev. {\bf 92}, 345 (1992);
          W.~B.~Davis, M.~A.~Ratner, M.~R.~Wasilewski, Chem. Phys. {\bf 281} 333 (2002).
          A.~S.~Lukas, P.~J.~Burchard, M.~R.Wasilewski, J. Chem. Phys. A {\bf 106} 2074 (2002).
                  M.~R. Wasielewski, in: {\it Photochemical Energy Conversion},
                  Eds. J.~Norris and  D.~Meisel (Elsevier, Amsterdam, 1989) p.~135;
                  M.~R. Wasielewski, {\it Tetrahedron} {\bf 45}, 4785 (1989).
\bibitem{mccusker00} A.~T.~Yeh, C.~V.~Shank, J.~K.~McCusker, Science {\bf 289} 935 (2000);
				D.~Kuciauskas, J.~E.~Monat, R.~Villahermosa, H.~B.~Gray, N.~S.~Lewis, J.~K.~McCusker, J. Phys. Chem. B {\bf 106} 9347 (2002).
\bibitem{Zare} R.~N.~Zare, Science {\bf 279} 1875 (1998).
\bibitem{Hochstrasser} M.~Lim and R.~M.~Hochstrasser, J.~Chem.~Phys. {\bf 115} 7629 (2001);
P.~Hamm, M.~Lim, W.~F.~DeGrado, and R.~M.~Hochstrasser, J. Chem. Phys. {\bf 112} 1907 (2000); 
M. T. Zanni, M. C. Asplund, and R. M. Hochstrasser, J. Chem. Phys. {\bf 114} 4579 (2001). 
\bibitem{textbook-spectroscopy} B. I. Stepanov and V. P. Gribkovskii,  {\it Theory of luminescence}; ed. S. Chomet (London, Iliffe, 1968).
\bibitem{Tannor-BOOK} D. J. Tannor, {\it Introduction to  Quantum Mechanics: A Time-Dependent Perspective}, (University Science Press, Sausalito, 2001).
\bibitem{non-lin-optics} S. Mukamel, {\it Principles of Nonlinear Optical Spectroscopy} (Oxford University Press, New York, 1995). 
\bibitem{tretiak_review} S. Tretiak, A. Saxena, R.~L.~Martin, and A.~R.~Bishop, J. Chem. Phys. {\bf 115} 699 (2001).
\bibitem{Takayoshi_ask_Piryatinski} S. Tanaka, S. Volkov, and S. Mukamel J. Chem. Phys. {\bf 118} 3965 (2003); S. Tanaka and S. Mukamel, ibid, {\bf 116} 1877 (2002)
\bibitem{Fayer} K.~A.~Merchant, D.~E.~Thompson, and M.~D.~Fayer, Phys. Rev. A {\bf 65} 023817 (2002); D. E. Thompson, K. A. Merchant, and M. D. Fayer, J. Chem. Phys. {\bf 115} 317 (2001).
\bibitem{Tokmakoff} O. Golonzka, M.~Khalil, N.~Demird\"oven, and A. Tokmakoff, J. Chem. Phys. {\bf 115} 10814 (2001); 
         M. Khalil, N. Demirdöven, and A. Tokmakoff, Phys. Rev. Lett. {\bf 90}, 047401 (2003);
         N. Demird\"oven, M. Khalil, and A. Tokmakoff, Phys. Rev. Lett. {\bf 89}, 237401 (2002). 
\bibitem{Jonas} J. D. Hybl, A.~A.~Ferro, and D.~M.~Jonas, J. Chem. Phys. {\bf 115} 6606 (2001).
\bibitem{sher91} N.~F.~Sherer, R.~Carlson, A.~Matro, M.~Du, A.~L.~Ruggiero, V.~Romero-Rochin, J.~A.~Cina, 
G.~R.~Fleming, and S.~A.~Rice, J. Chem. Phys. {\bf 95} 1487 (1991).
\bibitem{sher92} N.~F.~Sherer, A.~Matro, R.~J.~Carlson, M.~Du, L.~D.~Ziegler, J.~A.~Cina, and G.~R.~Fleming, 
J. Chem. Phys. {\bf 96} 4180 (1992).
\bibitem{Cina} J. A. Cina, D. S. Kilin, and T. S. Humble, J. Chem. Phys. {\bf 118}, 46 (2003). 
\bibitem{Shapiro} I. Sh. Averbukh, M. Shapiro, C. Leichtle and W. P. Schleich,
 	               Phys. Rev. A {\bf 59} 2163 (1999).
\bibitem{marc56} R.~A.~Marcus, J.\ Chem.\ Phys.\, {\bf 24}, 966 (1956);
                Rev.\ Mod. Phys.\ {\bf 65}, 599 (1993);          	
                R.~A.~Marcus and N.~Sutin, Biochim.\ Biophys.\ Acta {\bf 811}, 265 (1985).
\bibitem{foerster}Th. F\"orster,
 			 {\em Delocalized excitation and excitation transfer}
			 {\it in: Modern Quantum Chemistry,~Ed. O.~Sinanoglu}, (Academic, NY, 1965).
\bibitem{stepanov} M.V. Volkenshtein, M.A. Eliashevich, B.I. Stepanov, {\it Oscillations of Molecules}, (Moskow, 1947).
\bibitem{kuhn_may} V. May and O. K\"uhn, 
                  {\it Charge and Energy Transfer Dynamics in Molecular Systems}, (Wiley-VCH, 2000).
\bibitem{ross97} O. V. Prezhdo and P. J. Rossky, J. Chem. Phys. {\bf 107} 5963 (1997);
                 B.~J.~Schwartz, E~.R.~Bittner, O.~V.~Prezhdo, and P.~J.~Rosski, J.~Chem.~Phys. {\bf 109} 5942 (1996).
\bibitem{schatz-ratner-book} G. C. Schatz and M. A. Ratner, 
          {\it Quantum mechanics in chemistry}, (Englewood Cliffs, NJ, Prentice Hall, 1993).
\bibitem{Heller70} E.~J.~Heller, J. Chem. Phys. {\bf 62} 1544 (1975).
\bibitem{Mukamel89} Y.~J.~Yan and S. Mukamel, J. Chem.  Phys. {\bf 88} 5735 (1988); {\it ibid} {\bf 89} 5160 (1988);
		J.~Grad, Y.~J.~Yang, A.~Haque, and S. Mukamel, {\it ibid} {\bf 86} 3441 (1987);
		J.~Grad, Y.~J.~Yan, and S. Mukamel, Chem. Phys. Lett. {\bf 143} 291 (1987);
		S. Mukamel, J. Phys. Chem. {\bf 88} 3185 (1984).         
\bibitem{bill_miller} W. H. Miller, J. Chem. Phys. {\bf 62} 1899 (1975).
\bibitem{feynman} R.~P.~Feynmann, {\it Quantum Mechanics and Path Integrals} (NY, McGrow-Hill, 1965).
\bibitem{tull98} J.~C.~Tully, in {\it Classical and Quantum Dynamics in Condenced Phase Simulations},
                 ed. B.~J.~Berne, G.~Cicotty, and D.~F.~Coker (World Scientific, 1998).
\bibitem{coke93} D.~F.~Coker, in {\it Computer Simulations in Classical Physics}, ed. 
                 M.~P.~Allen and D.~J.~Tildesley (Kluwer, 1993).
\bibitem{schw03} C.~J.~Smallwood, W.~B.~Bosma, R.~E.~Larsen, and B.~J.~Schwartz, J. Chem. Phys. {\bf 119} 11263 (2003); R.~E.~Larsen and B.~J.~Schwartz, 
                 {\it ibid} {\bf 119} 7672 (2003).
\bibitem{referees} S.~Karabunarliev and E.~R.~Bittner, J.~Chem.~Phys. {\bf 119} 3988 (2003).
\bibitem{pere00} O. V. Prezhdo and Yu. V. Pereversev, J. Chem. Phys. {\bf 113} 6557 (2000).
\bibitem{Prezhdo2002} O. V. Prezhdo, J. Chem. Phys. {\bf 117} 2995 (2002).
\bibitem{pere02} O.~V.~Prezhdo and Yu. V. Pereverzev, J. Chem. Phys. {\bf 116} 4450 (2002).
\bibitem{pahl02} E.~Pahl and O.~V.~Prezhdo, J. Chem. Phys. {\bf 116} 8704 (2002).
\bibitem{broo03} C.~Brooksby, O.~V.~Prezhdo, Chem. Phys. Lett. {\bf 378} 533 (2003); J. Mol. Struct. (Theochem) {\bf 630} 45 (2003).
\bibitem{Wick1950} G.~C.~Wick, Phys. Rev. {\bf 80} 268 (1950);
			R.~P.~Feynman, Phys. Rev. {\bf 76} 749 (1949);
			F.~J.~Dyson, Phys. Rev. {\bf 75} 286 (1949).
\bibitem{Leggett} A. Leggett, Rev. Mod. Phys. {\bf 59} 1 (1987).
\bibitem{general_QO} P. Meystre and M. Sargent III, 
                     {\it Elements of Quantum Optics} (Springer, 1990).
			K. Blum, {\it Density Matrix, Theory and Applications} (Plenum, 1981);
		Y. Yamamoto, {\it Mesoscopis Quantum Optics} (Wiley, 1999);
R. Puri, {\it Mathematical Methods in Quantum Optics} (Springer, 2000);
S. M. Barnett and P. M. Radmore, {\it Methods in Theoretical Quantum Optics} (Clarendon, 1997);
\bibitem{Allen_Eberly} L. Allen and J. H. Eberly, {\it Optical Resonance and Two-Level Atoms} (Wiley-VCH, 1975). 
\bibitem{Jaynes-Cummings63} E.~T.~Jaynes and F.~W.~Cummings, 
Proc. IEEE {\bf 51} 89 (1963)
\bibitem{KnightJCM} B.~W.~Shore and P.~L.~Knight, J. Mod. Optics {\rm 40} 1195 (1993).
\bibitem{modernALGEBRAjcm} D. Bonatosos, Phys. Rev. A {\bf 47}, 3448 (1993); 
				     S. Yu et.al., Phys. Rev. A {\bf 52}, 2585 (1995);
			N. Alvarez M. and V. Hussin, J. Math. Phys. {\bf 43}, 2063 (2002). 
\bibitem{Ackerhart} J.~R.~Ackerhalt and K.~Rzazewski, Phys. Rev. A {\bf 12} 2549 (1975).
\bibitem{Scully_BOOK} M. O. Scully and M. S. Zubairu, {\it Quantum Optics}, (Cambridge, 1997).
\bibitem{cumi65} F.~W.~Cummings, Phys. Rev. {\bf 140} A1051 (1965).
\bibitem{naro80} J.~H.~Eberly, N.~B.~Narozhny, and J.~J.~Sanchez-Mondragon, Phys. Rev. Lett. {\bf 44} 1323 (1980).
\bibitem{Berman_Gena_BOOK} G.~P.~Berman, E.~N.~Bulgakov, and D.~D.~Holm, {\it Crossover-Time in Quantum Boson and Spin Systems} (Springer, 1994).
\bibitem{Carmichael_BOOK} H.~J.~Carmichael, {\it Statistical Methods in Quantum Optics 1, Master Equations and Fokker-Plank Equations} (Springer, 1999). 
\bibitem{redf55} A.~G.~Redfield, Phys. Rev. {\bf 98}, 1787 (1955);
                 IBM J. Res. Dev. {\bf 1}, 19 (1957);
                 Adv. Magn. Reson. {\bf 1}, 1 (1965). 
\bibitem{kilin2000} D. Kilin, M. Schreiber, J.  Lumin. {\bf 92}, 13 (2001).
\bibitem{poll94} W.~T.~Pollard and R.~A.~Friesner,
                 J.~Chem.~Phys. {\bf 100}, 5054 (1994). 
\bibitem{kenkre95} M. I. Sakola, A. R. Bishop, V. M. Kenkre, and S. Raghavan, Phys. Rev. B {\bf 52} R3824 ('95).  
\bibitem{S_Ya_Kilin_review_1987} P.~A.~Apanasevich, S.~Ya.~Kilin, and A.~P.~Nisovtsev, 
        J. Appl. Spectr., {\bf 47} 1213 (1987); 
        S.~Ya.~Kilin and T.~B.~Krinitskaya, Phys. Rev. A {\bf 48} 3879 (1993); 
        D. Mogilevtsev and S. Kilin, Phys. Rev. A 67, 023815 (2003).  
\bibitem{hols59} T.~Holstein, Ann. Phys. {\bf 8} 325 (1959).
\bibitem{Freming3PEPS} R. Agarwal, B.~S.~Prall, A.~H.~Rizvi, M.~Yang, and G.~R.~Fleming, J. Chem. Phys. {\bf 116} 6243 (2002).
\end{thebibliography}

\bibliographystyle{prsty}
\newpage
\newpage
\begin{table}[ht!]
\begin{center}
\caption[electronic and vibrational pure quantum energy]
{\small  Electronic and vibrational energies partitioned
into quantum and classical contributions. The energies 
are given in the ladder, phase space and dispersion operator
representations.
In both electronic and vibrational subsystems the energy flows between the quantum and quasi-classical contributions.}
\end{center}
\label{TAB_pure_quantum}
\begin{tabular}{cccc} \tiny
            & \multicolumn{3}{c}{Energy}   \\ \hline
            &     total          &    quasiclassical &    pure quantum \\ \hline \\
& \multicolumn{3}{c}{Vibrational} \\ \hline

ladder   operators $a^+$, $a$                  & $\langle a^+a\rangle+1/2$ & $\langle a^+\rangle   \langle a \rangle$       &       $\langle a^+a \rangle -\langle a^+ \rangle \langle a \rangle $\\
phase space operators $p$, $q$                  & $\frac{\langle p^2\rangle}{2m} + \frac{m\omega^2\langle q^2 \rangle}{2}$    & $\frac{\langle p\rangle^2}{2m} + \frac{m\omega^2\langle q \rangle^2}{2}$    & $\frac{\langle p^2\rangle- \langle p \rangle^2 }{2m} + \frac{m\omega^2\left(\langle q^2 \rangle-\langle q \rangle^2 \right)}{2}$   \\
dispersions                                    & --- & --- &  $\frac{1}{2m}\sigma_q+ \frac{m\omega^2}{2}\sigma_p$\\
equation                                       & (\ref{EQ_vib_total})      & (\ref{EQ_vib_class})                           &     (\ref{EQ_disper_bosonic}), (\ref{EQ_sum_disp})  \\ 

\\
&\multicolumn{3}{c}{Electronic} \\ \hline
ladder operators, $S_+$, $S_-$                 &  $\langle S_+S_-\rangle$    & $\langle S_+\rangle   \langle S_- \rangle$       &       $\langle S_+S_- \rangle -\langle S_+ \rangle \langle S_- \rangle $\\
projection operators, $S_\x$, $S_\y$, $S_\z$   & $\langle S_\z \rangle +1/2$ & $\langle S_\x\rangle^2 + \langle S_\y \rangle^2$ & $\langle S_\z \rangle +1/2  - \left( \langle S_\x\rangle^2 + \langle S_\y \rangle^2  \right)$ \\
dispersions                                    & --- & $\frac12-\left(\sigma_{S_\x} + \sigma_{S_\y} \right) $ &    $\langle S_\z \rangle +\left(\sigma_{S_\x} + \sigma_{S_\y} \right)$    \\
equation                                       & (\ref{EQ_el_total})  & (\ref{EQ_short_el_sum})  & (\ref{EQ_fermionic_uncertainty})  \\
\end{tabular}
\end{table}

\newpage

\begin{table}[ht!]
{\rotate{\begin{minipage}{22cm}
\begin{center}
\caption[Key characteristics of electronic subsystem]
{\small Key characteristics of the electronic subsystem: 
Bloch vector squared ($R^2$), pure quantum energy 
($E_{\rm el}^{\rm quant}$), and 
fidelity (Tr$\rho^2$) are represented
by the expectation values of 
the ladder, projection and dispersion operators.
These characteristics obey the equalities 
for pure states and the inequalities for mixed states.}
\end{center}
\label{TAB_mixed_state}
\begin{tabular}{cccc} \footnotesize 
                                               &  $R^2 \le 1/4$                              &   $E_{\rm el}^{\rm quant}\ge 0$                  & ${\rm Tr}\rho_{\rm}^2 \le 1$ \\ \\ \hline \\
ladder, $S_+$, $S_-$                 &  $1/4-\langle S_+S_-\rangle \langle S_-S_+\rangle + \langle S_+\rangle \langle S_-\rangle $ 
                                                                                             & $\langle S_+ S_-\rangle -  \langle S_+ \rangle \langle S_-\rangle$       
                                                                                                                                                &       $1+2\langle S_+\rangle \langle S_- \rangle -2 \langle S_+ S_- \rangle \langle S_- S_+ \rangle $\\ \\
projections, $S_\x$, $S_\y$, $S_\z$   & $\langle S_\x \rangle^2+\langle S_\y \rangle^2+\langle S_\z \rangle^2 $                 
                                                                                             & $\langle S_z\rangle +1/2 -\langle S_\x\rangle^2 - \langle S_\y \rangle^2$ 
                                                                                                                                                & $1/2-2\langle S_\z \rangle^2 +2 \left( \langle S_\x\rangle^2 + \langle S_\y \rangle^2  \right)$ \\ \\
dispersions                                    & $1/2+\langle S_z \rangle^2-\sigma_{S_\x}-\sigma_{S_\y}$ 
                                                                                             & $\langle S_\z \rangle +\sigma_{S_\x} + \sigma_{S_\y}$ 
                                                                                                                                                & $1/2 -2\langle S_\z \rangle^2 +1-2\left(\sigma_{S_\x} + \sigma_{S_\y} \right)$    \\
\\
\end{tabular}
\end{minipage}}}
\end{table}

\normalsize
\newpage
\vspace{3cm}    

\newpage

\section*{Figure Captions}

{\bf FIG. 1} Electronic states 
and state-specific potential energy curves as functions of 
the vibrational coordinate of the dimer described by 
the Hamiltonian~(\ref{EQ_convenient_ham}),
appearing in the theory of molecular excitons and
also known as the Marcus model.
The
electronic states are separated by the energy gap $\varepsilon=\hbar\Omega$.
The equilibrium positions in the vibrational coordinates of the two
electronic states are displaced by the
value specified in 
Eq.~(\ref{EQ_coordinate_shift}) and related
to the nuclear reorganization $(\bar\alpha+\bar\alpha^*)/2$.

\vspace{.2cm}

{\bf FIG. 2} 
The expectation value 
of the population inversion operator $S_z$ 
as a function of time 
computed exactly, solid line,
and approximately
by Eq.~(\ref{EQ_solution}), dotted line.
{\bf (a)} illustrates the quasi-classical
case of a weak electron-phonon
coupling $g=0.025$ and a large amplitude 
vibrational motion $\langle a^+a\rangle(t=0)=49$.
{\bf (b)} corresponds to a deep quantum
case of a strong coupling
$g=0.25$ and a smaller amplitude vibration
$\langle a^+a\rangle(t=0)=9$.
$\left<S_\z\right>_{t=0}=-1/2$ and $\delta=0$
in both cases.
Good agreement
between the exact and approximate solutions 
is observed during the first several periods 
of the population inversion. For both types
of initial conditions, the approximate
solution correctly reproduces the dephasing envelope 
describing the decay in the amplitude 
of the oscillation of the electronic populations.

\vspace{.2cm}

{\bf FIG. 3}
Population inversion $t^{-1}_+$ and relaxation 
$t^{-1}_-$ rates, Eq.~(\ref{EQ_inv_rel}), 
labeled by ''$+$'' and ''$-$'',
as functions of  {\bf (a)} detuning $\delta$,
and {\bf (b)} amplitude $\bar \alpha$,
Eq.~(\ref{EQ_coordinate_shift}).
The remaining parameters
are kept constant at quasi-classical
$\langle a^+a\rangle=49$, $g=0.025$ (solid line)
and quantum
$\langle a^+a\rangle=9$, $g=0.25$ (dashes) 
values used in Fig.~2. 
The dependence of the inversion relaxation rate
on the amplitude of vibrational motion displays branching in (b).
The inversion $t_+^{-1}$ and relaxation $t_-^{-1}$ rates
coincide for small values 
of the vibrational amplitudes $\bar \alpha$.
The branches split at $\bar \alpha =1$.

\vspace{.2cm}

{\bf FIG. 4}
Evolution of the sum of the population 
inversion $2\langle S_\z\rangle$ and 
the electron-phonon correlation 
$\bar\alpha^{-1}\langle\beta\rangle$.
The purely real $2\langle S_\z\rangle$ and imaginary
$\bar\alpha^{-1}\langle\beta\rangle$ 
functions
shown in  {\bf (a)} 
are dynamic conjugates
in the sense of Eq.~(\ref{EQ_blocks}) with $n=0$.
The absolute value
$\left| \langle 2S_\z + \bar\alpha^{-1}\beta\rangle \right|$ 
is shown in  {\bf (b)}.
The initial conditions and parameters are
$\left<a^+a\right>_{t=0}=9$, $\left<S_\z\right>_{t=0}=-1/2$, 
$g=0.25$, $\delta=0$,  corresponding to Fig.~2(b).
The population inversion $S_\z$ 
and 
the electron-phonon
correlation 
$\beta$
oscillate and dephase with the same rate.
The amplitude $\left| \langle 2S_\z + \bar\alpha^{-1}\beta\rangle \right|$ 
displays smooth dephasing dynamics 
without the quick oscillations.

\vspace{.2cm}
{\bf FIG. 5}  Key characteristics of the vibrational
subsystem.
{\bf (a)}:
Dispersions $\sigma_q$, $\sigma_p$ 
of the vibrational coordinate $q$
and momentum $p$ are shown 
by the solid and dotted lines, 
Eqs.~(\ref{EQ_vibrational_dispersion})
and~(\ref{EQ_disp_momentum}), respectively.
The sum $\sigma_q+\sigma_p$  Eq.~(\ref{EQ_sum_disp}), 
shown 
by the dashed line starts from the minimum uncertainty width
and smoothly increases
to its maximum value that is equal to 
the total energy of the system.
The sum of dispersions serves as an envelope 
for the individual dispersions.
The slow increase of the dispersion is due to 
the splitting of the
original Gaussian wavepacket into two Gaussians 
that oscillate
independently and pass through each other leading to
the rapid oscillations of $\sigma_q$ and $\sigma_p$.
{\bf (b)}:
Shown are the total energy of the vibrational mode 
Eq.~(\ref{EQ_vib_total}), solid line, together with its
quasi-classical contribution, Eq.~(\ref{EQ_vib_class}), dots,
and the purely quantum contribution, Eq.~(\ref{EQ_sum_disp}), 
dashes. 
The vibrational energy flows between its
quasiclassical and purely quantum contributions.
At $t=0$ the quantum contribution equals 
zero-point energy, and the rest of the vibrational
energy is quasiclassical.
When the total vibrational energy stops oscillating,
it is purely quantum.
When the oscillation resumes, the vibrational energy is
dominated by the classical contribution.
\comment{
Interesting to note, that the portion of
classical contribution becomes less and less for each cycle.
The system approaches the essentially quantum state.
}
The sum of dispersions in (a)
is an alternative representation of the quantum energy 
in (b), dashed lines.
The initial conditions and model parameters are the same
as in Fig.~2(b).
Note, that the phase of the vibrational energy oscillation
in this figure is opposite to the phase of the electronic
population in Fig.~2(b), due to conservation
of the total electron-vibrational energy.

\vspace{.2cm}
{\bf FIG. 6} {\bf (a)}  Time slices of the vibrational wavepacket in 
coordinate representation, 
{\bf (b)} the expectation value of the vibrational coordinate $q$, 
 {\bf (c)}
and the expectation value of the population 
inversion $\langle S_\z \rangle +1/2$.
The initial conditions and model parameters are 
taken from Ref.~[59].
Transfer of population between
the two states affects the wavepacket dynamics. The original Gaussian wavepacket
splits into two wavepackets correlated with each electronic 
state. As a result of the splitting
the oscillation of the average vibrational coordinate decays. Most interestingly,
the oscillation of the electronic population 
decays as well. 
The decay in the oscillation of the electronic population,
also known as dephasing, typically occurs by coupling 
to a heat bath. Remarkably, a single vibrational mode 
acts as a quasi-heat bath and induces population
dephasing over a finite time.

\vspace{.2cm}
{\bf FIG. 7}  Key characteristics of the electronic 
subsystem, cf. Fig.~6 
 for the vibrational subsystem.
{\bf (a)}:
Dispersions $\sigma_{S_\x}$, $\sigma_{S_\y}$ of the
conjugate electronic operators $S_\x$ and $S_\y$
are shown by the solid and
dashed lines, Eqs.~(\ref{EQ_short_disp}) 
and~(\ref{EQ_short_disp_y}), respectively.
The sum $\sigma_{S_\x}+\sigma_{S_\y}-1/4$, 
Eq.~(\ref{EQ_short_el_sum}) 
shown by the thick dot-dashed line
starts from the maximum value of $1/4$, corresponding to a pure
state, and varies more smoothly than the components 
$\sigma_{S_\x}$, $\sigma_{S_\y}$, creating an envelope.
In contrast to
the vibrational dispersions, Fig.~5, 
which increase
 due to the electron-vibrational interaction,
the electronic dispersions decrease.
{\bf (b):} 
Shown are the 
total energy of 
the electronic subsystem  $\langle  S_+ S_-\rangle$, 
Eq.~(\ref{EQ_el_total}), solid line, 
its quasiclassical contribution, Eq.~(\ref{EQ_el_classic}),
dots, and the 
purely quantum contribution, Eq.~(\ref{EQ_el_quantum}), dashes.
The electronic and vibrational energies, this figure and
Fig.~5,
oscillate with opposite phases so that the total electron-vibrational
energy remains constant.
The electronic energy flows 
between its quasiclassical and purely quantum contributions.
At $t=0$ both quantum and classical contributions equal zero 
since system is prepared in  the state $\langle S_\z \rangle(0) =-1/2$
During the silent period, 
when the population inversion stops oscillating,
the electronic energy contains the largest portion of the classical contribution, 
while the vibrational energy is purely quantum.
The sum of dispersions in (a) 
 is an alternative representation of the classical
component of the electronic energy in (b), 
up to the linear transformation of Eq.~(\ref{EQ_short_el_sum}).
The initial conditions 
and model parameters are the same as in Fig.~2(b).

\vspace{.2cm}
{\bf FIG. 8} Slowly oscillating
combinations of the expectation values.
The initial conditions and model parameters 
are the same as in Fig.~2(b).
{\bf (a)}.  
Time dependence of the sums of the electronic and
vibrational dispersions, solid and dashed 
lines, respectively.
Surprisingly, the sum of the electronic dispersions, 
Eq.~(\ref{EQ_short_el_sum}), gives 
the quasiclassical contribution to the electronic energy,
Eq.~(\ref{EQ_el_classic}),
 while
the sum of vibrational dispersions
gives the pure quantum contribution to 
the vibrational energy, Eq.~(\ref{EQ_sum_disp}).
{\bf (b):} 
Imaginary part of the correlation between the 
electronic and vibrational subsystems.
The expectation value of the correlator 
Eq.~(\ref{EQ_correlator})  
is purely imaginary, its
magnitude is displayed on the $y$-axis.  
The electron-vibrational correlator starts at 0 and
evolves on the slow time-scale of formation and 
dephasing of the electron-vibrational entanglement.  
The extrema of the sums of the electronic and
vibrational dispersions in (a)
correspond to the fastest change in the correlation
in (b) and vice versa.
The characteristics displaced in the figure are
expressed in several ways in Table~I.

%
%

{\bf FIG. 9} 
Trajectories of the Bloch vector, Bloch vector squared,
and a relationship between the key
characteristics of the electronic subsystem.
{\bf (a):}
Bloch vector, Eq.~(\ref{EQ_bloch}),
as a function of time during the first one and a half 
population inversions,
for the same initial conditions and model parameters as in 
Fig.~2(b).
The Bloch vector is defined by the 
three spin projections of the electronic two-level system 
$\langle S_\x \rangle$,
$\langle S_\y \rangle$,
$\langle S_\z \rangle$ and is shown together with 
a Bloch sphere of radius 1/2.
At $t=0$ the Bloch vector
connects the origin and the south
pole of the Bloch sphere.  
The Bloch vector revolves from the south pole to
the north pole and changes in length.
{\bf (b):}
The squared length of the Bloch vector as a function of time.
$R^2=1/4$ at $t=0$, decreases at $0<\omega t< 2$,
and rises to a local maximum at $\omega t \simeq 6$.
 {\bf (c):}
Key characteristics of the electronic subsystem
as functions of the horizontal projection of 
the Bloch vector, 
Eq.~(\ref{EQ_horisontal}),
for $\langle S_z \rangle=0$,
$\langle S_+S_-\rangle =\langle S_-S_+\rangle = 1/2$.
Shown are the Bloch vector length squared
Eq.~(\ref{EQ_Bloch_ladder}), solid line,
the sum of dispersions for the electronic subsystem   
Eq.~(\ref{EQ_short_el_sum}), dot-dashes,
the quantum contribution to the electronic energy 
Eq.~(\ref{EQ_el_quantum}), dots, and fidelity
Eq.~(\ref{EQ_fidelity}), dashes.
The right edge, 
$\langle S_\x \rangle^2 + \langle S_\y \rangle^2=0.25$,
corresponds to pure states characterized by maximal fidelity,
${\rm Tr}\rho^2_{\rm el}=1$ and $R^2=1/4$.
The left edge,
$\langle S_\x \rangle^2 + \langle S_\y \rangle^2=0$,
describes the mixed state with minimal fidelity,  
${\rm Tr}\rho^2_{\rm el}=1/2$ and $R^2=0$.
Interaction with the bosonic mode drives 
the electronic mode from a pure to a mixed state.

\newpage

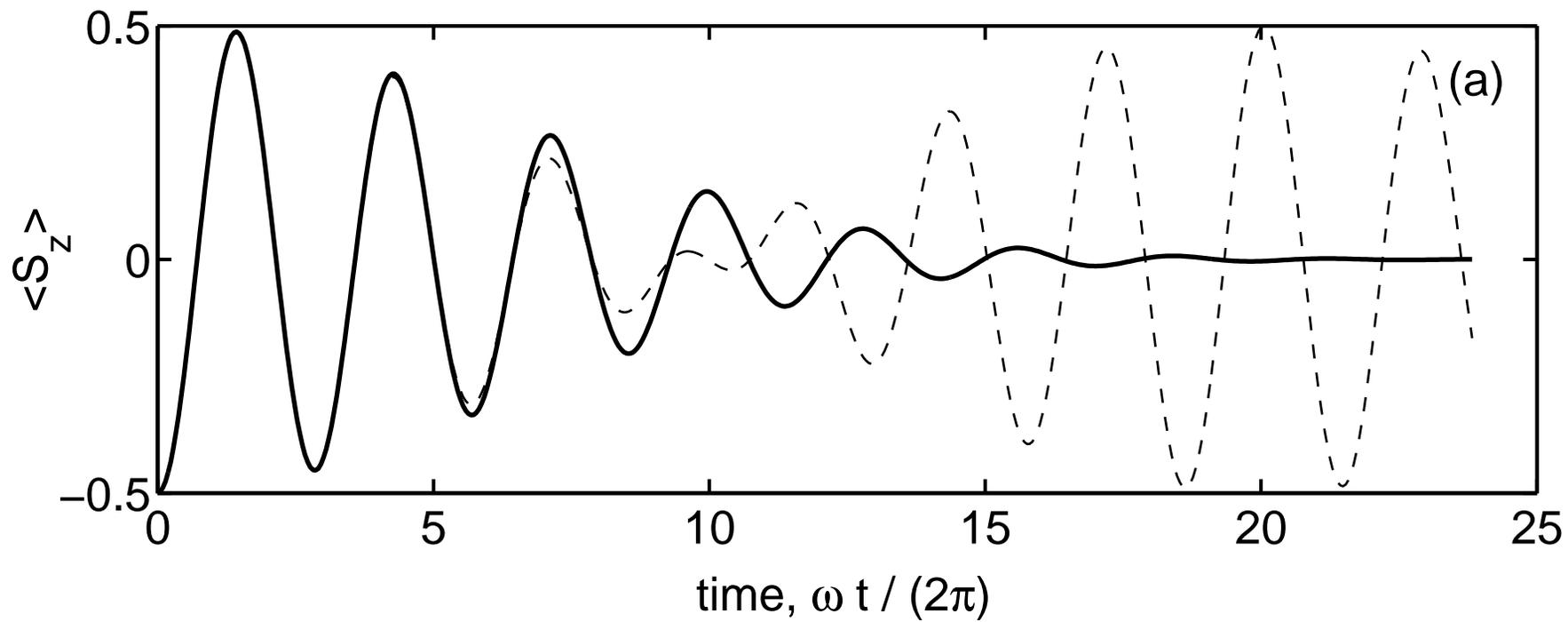
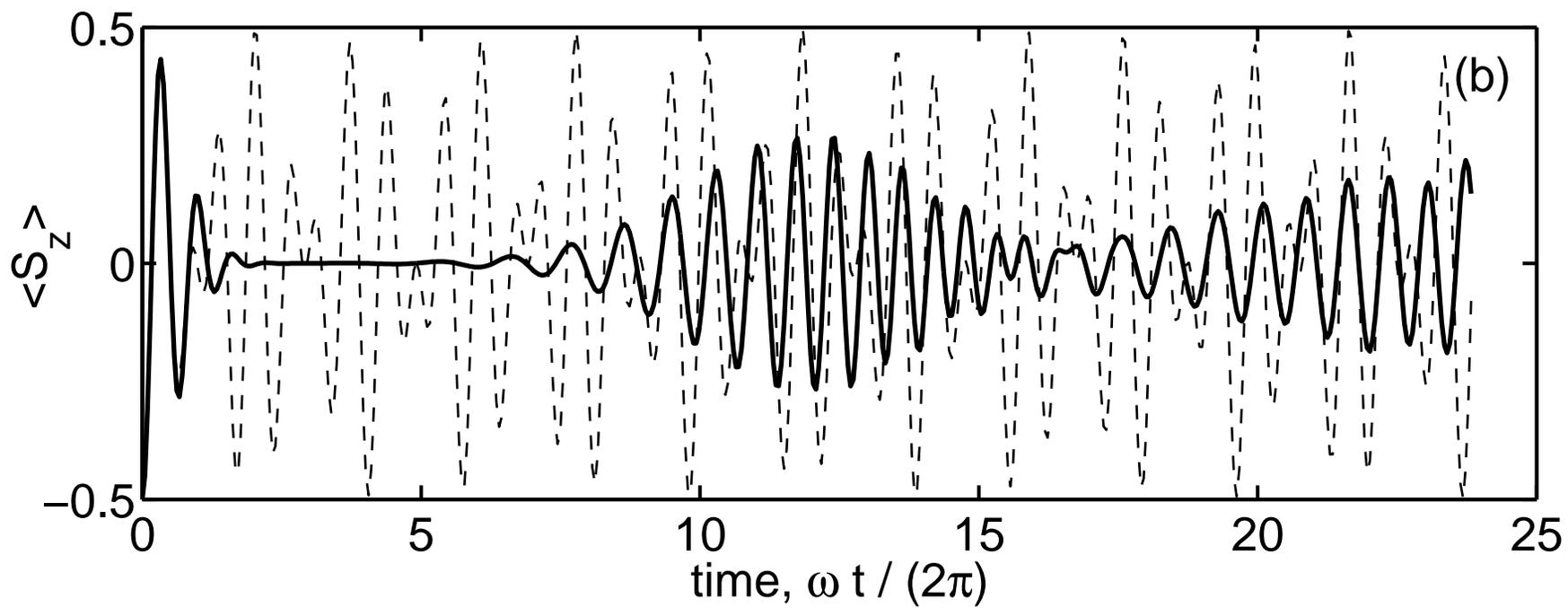
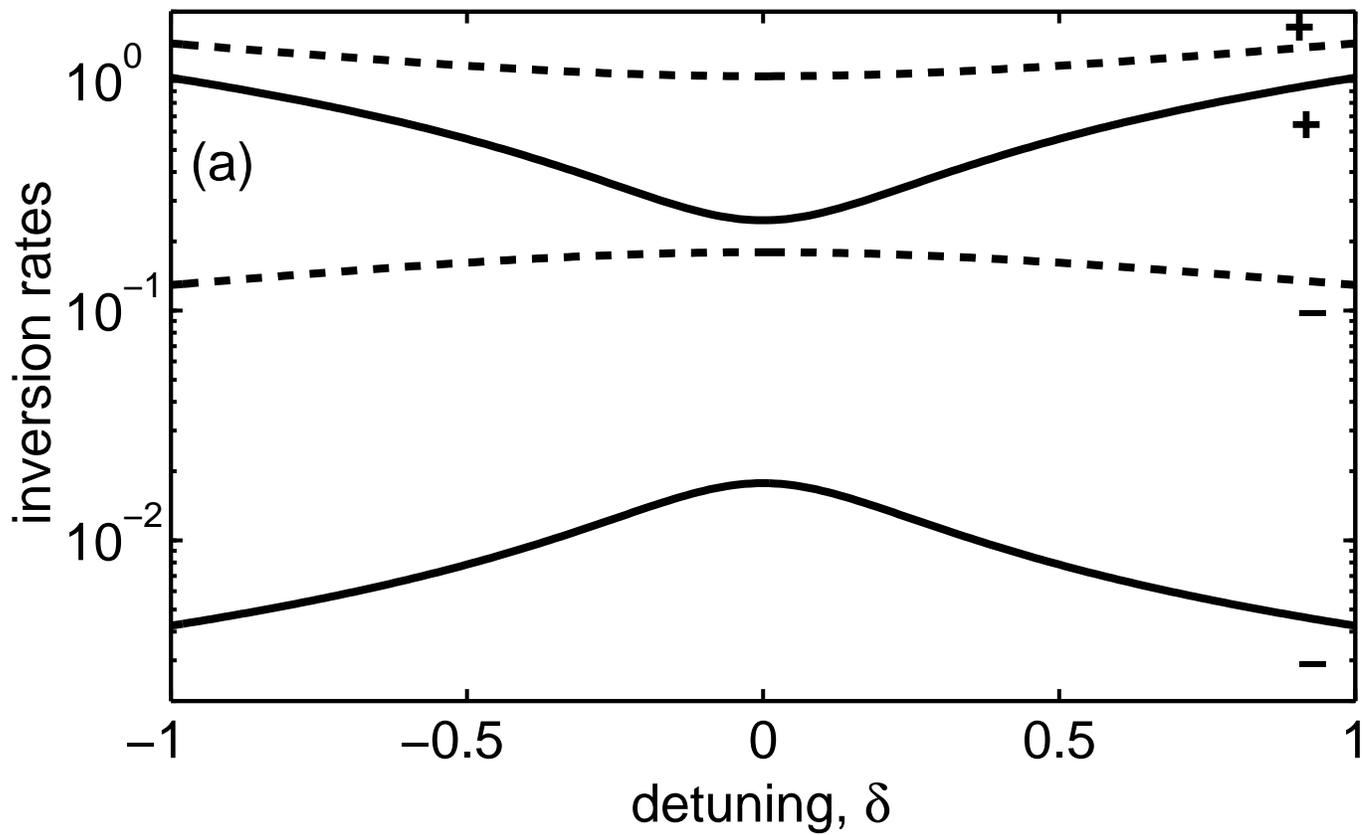
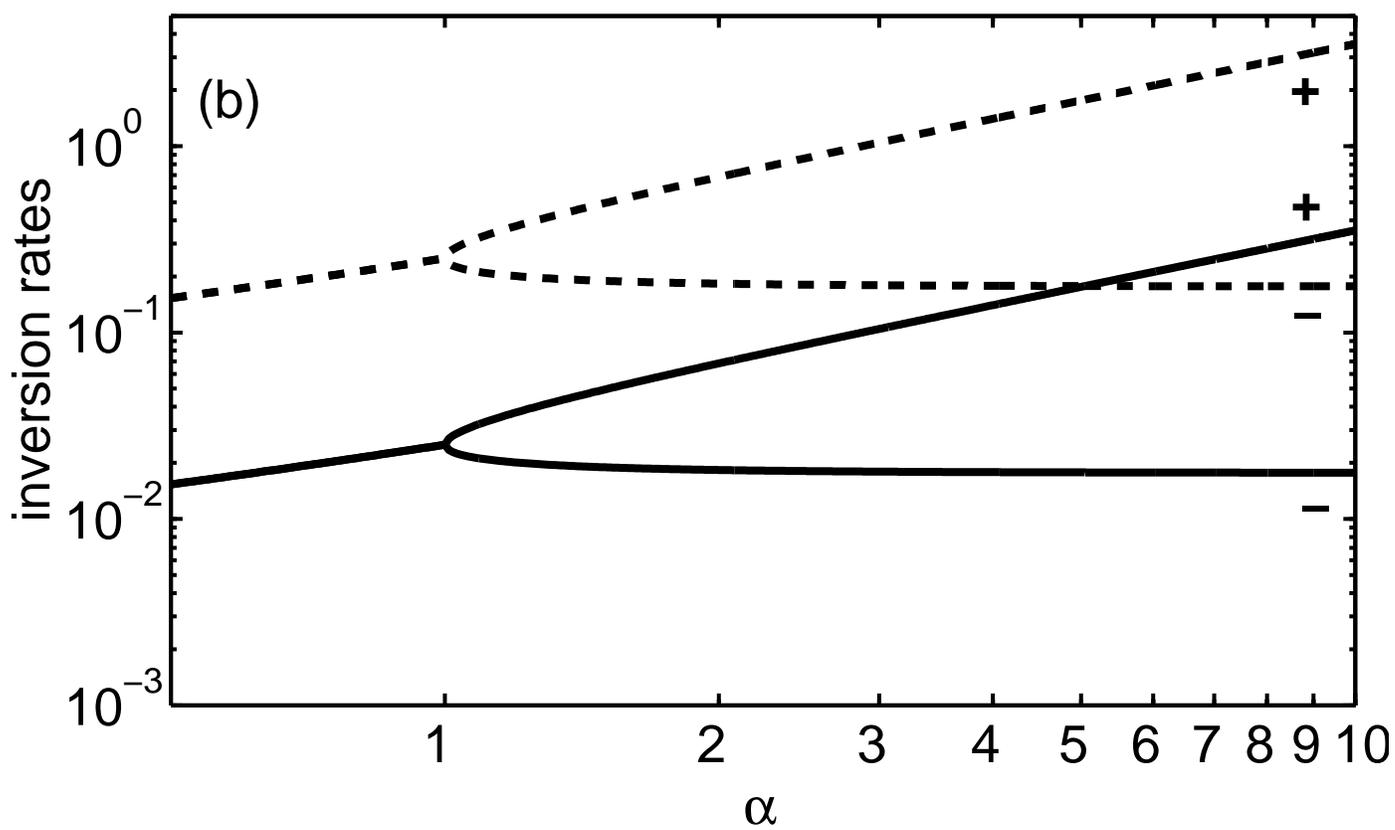
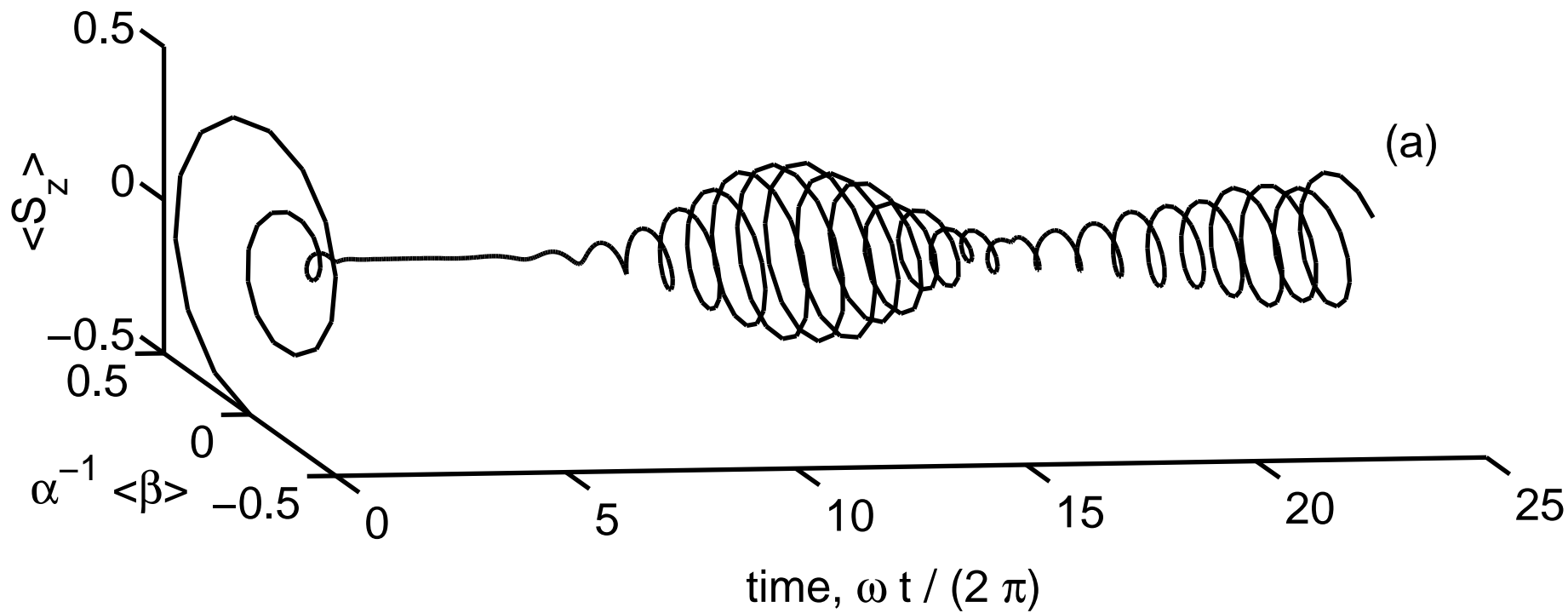
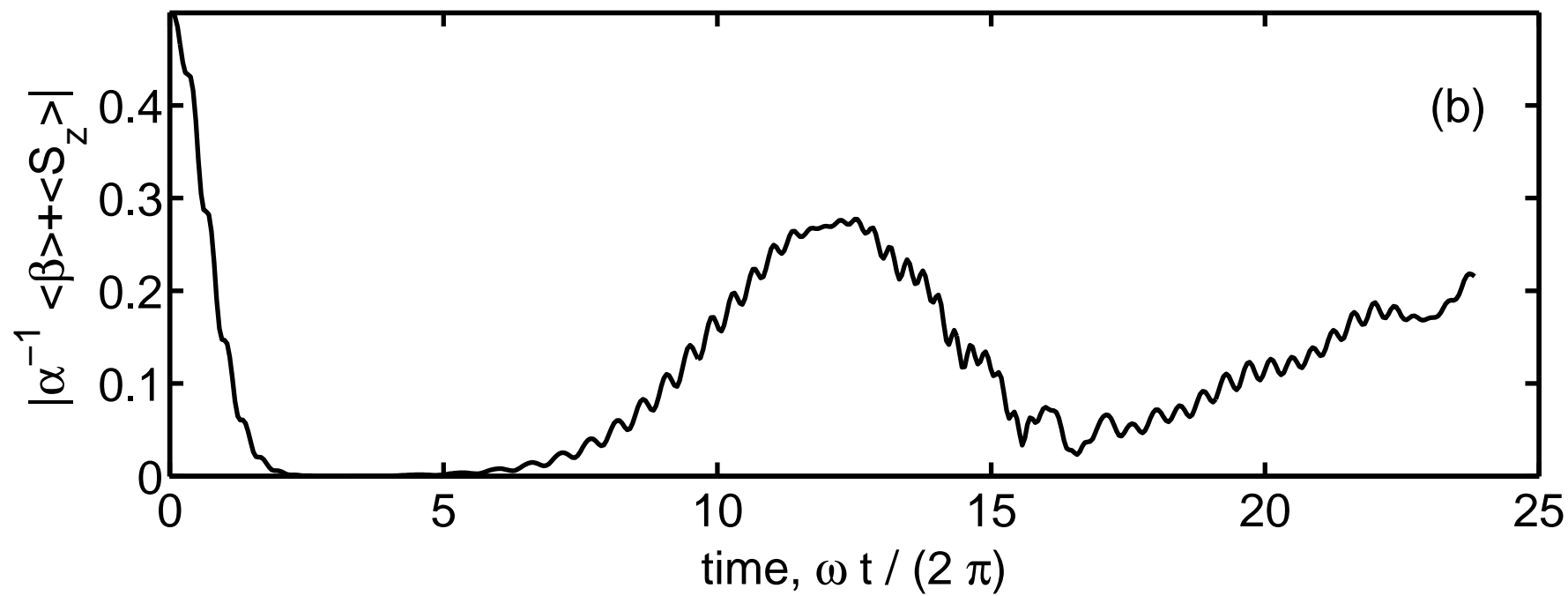
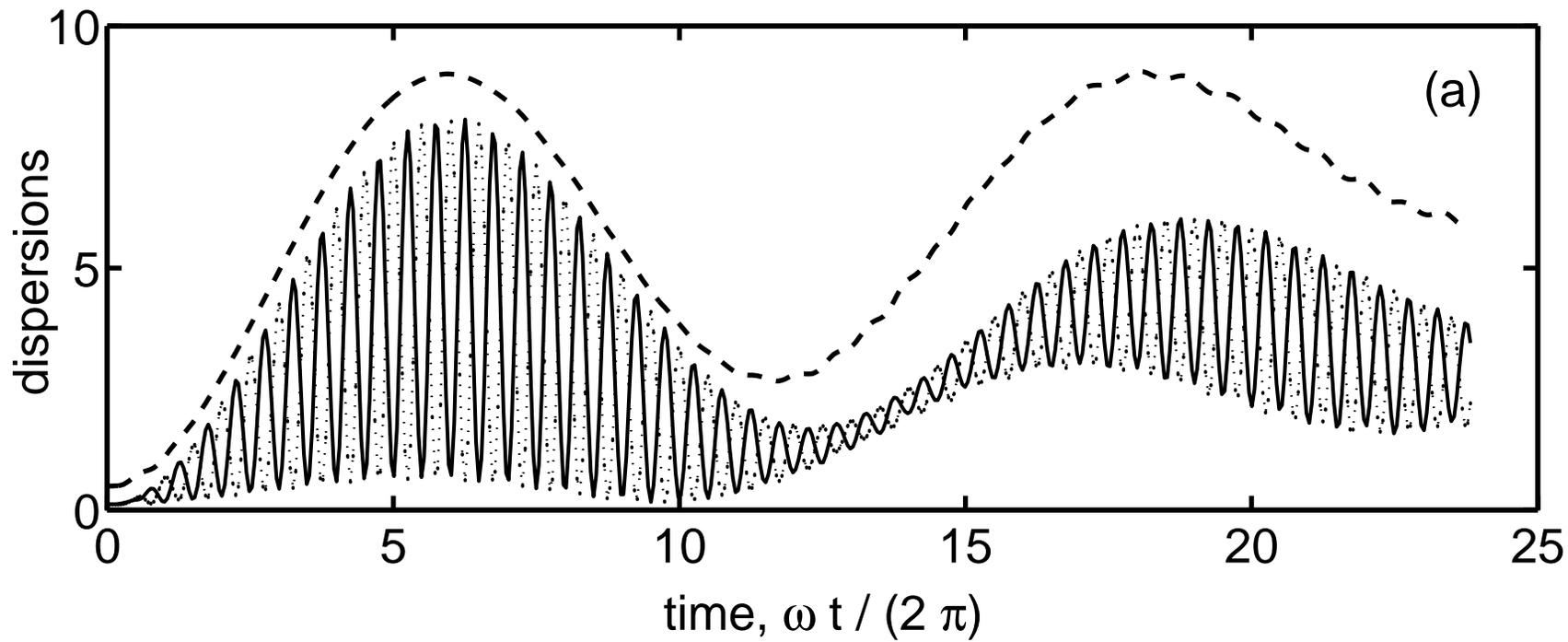
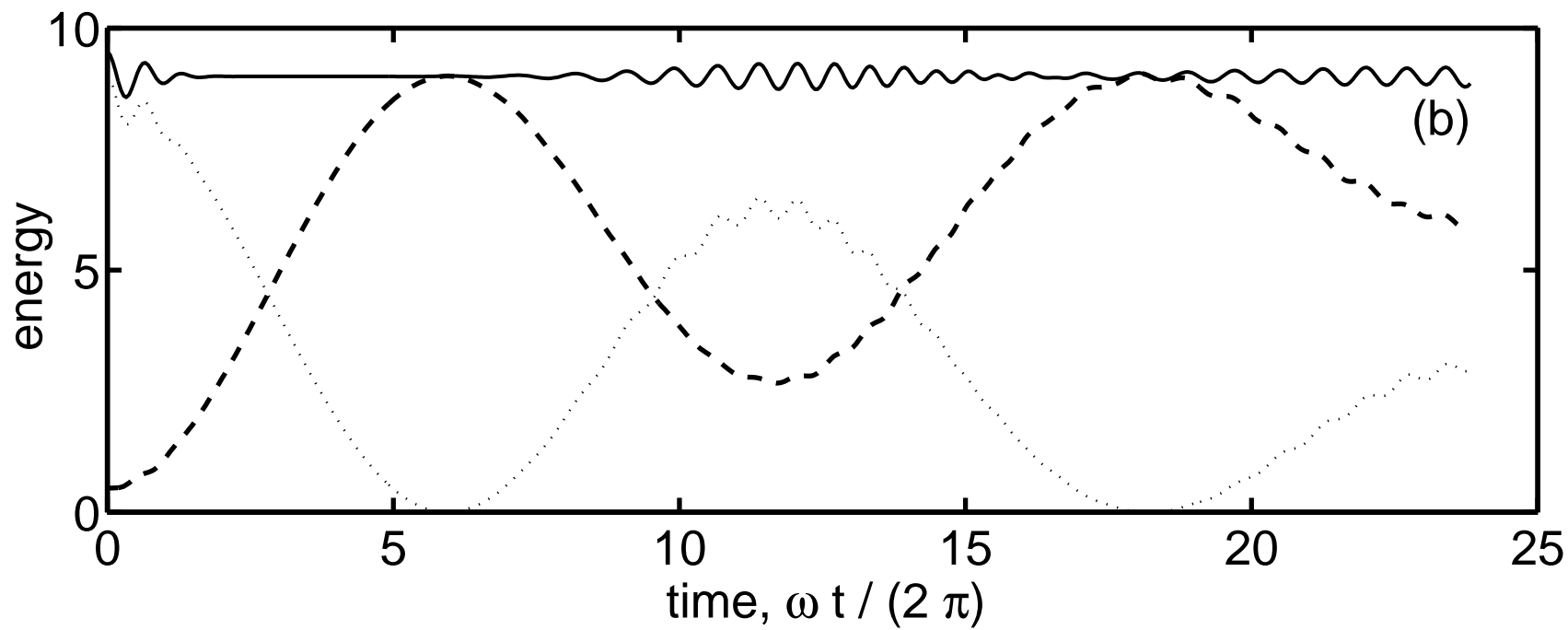
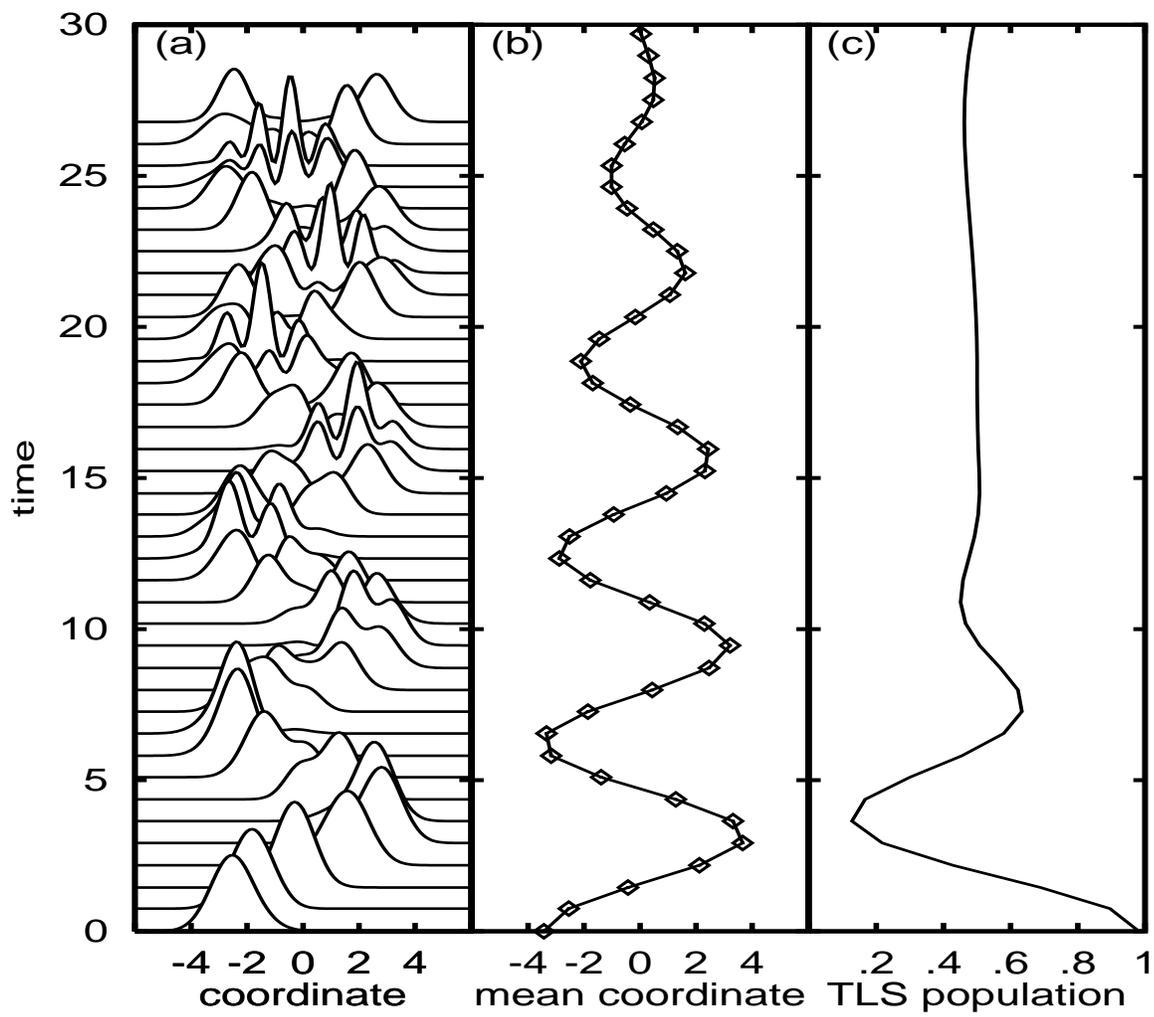
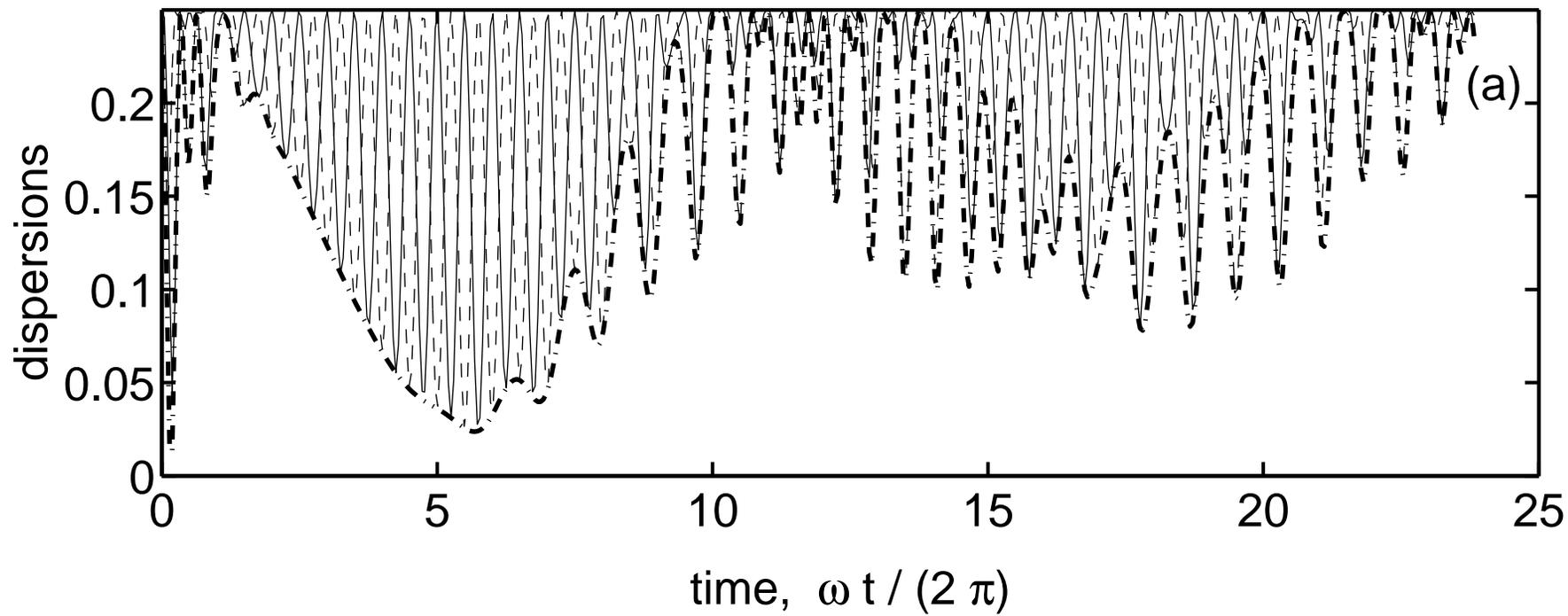
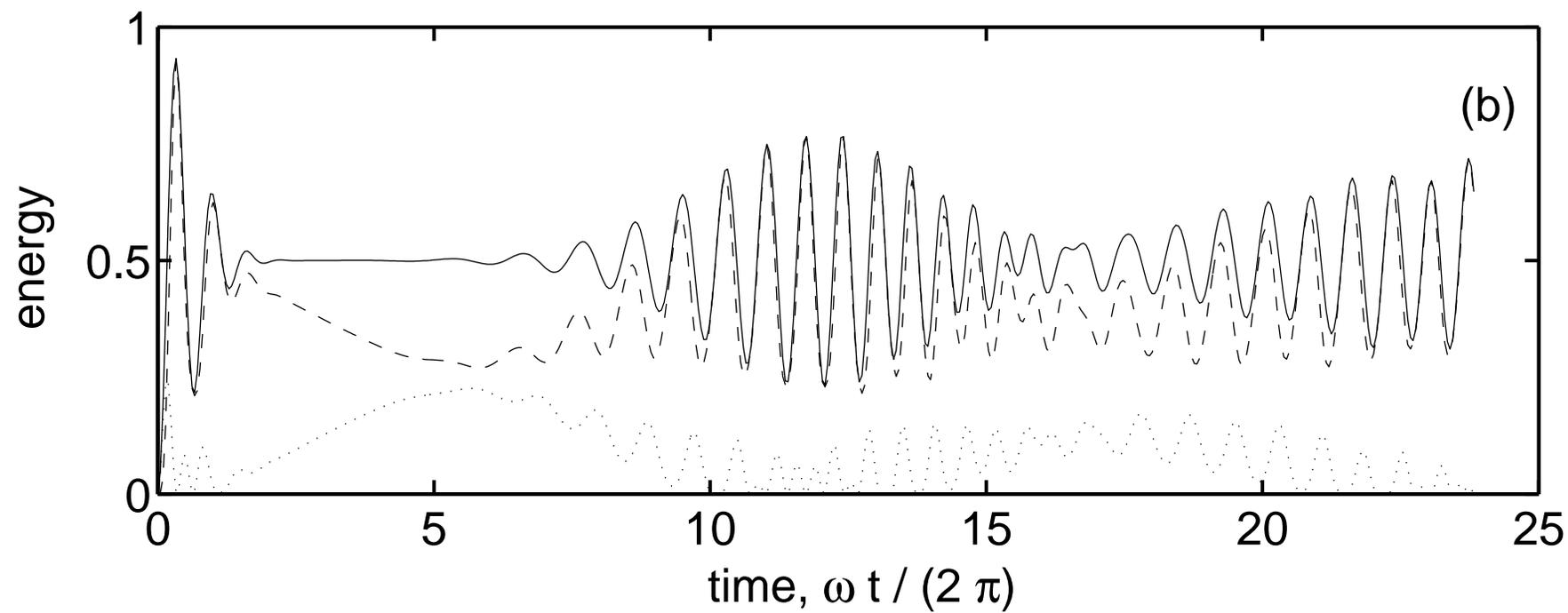
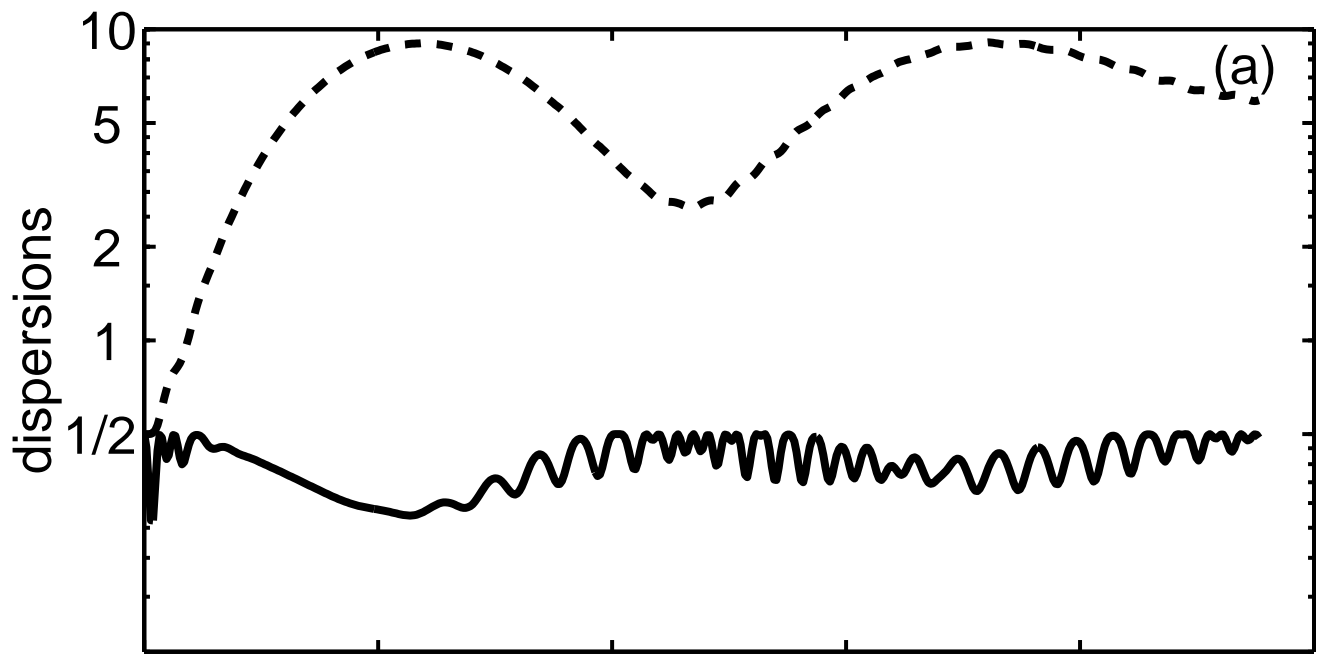
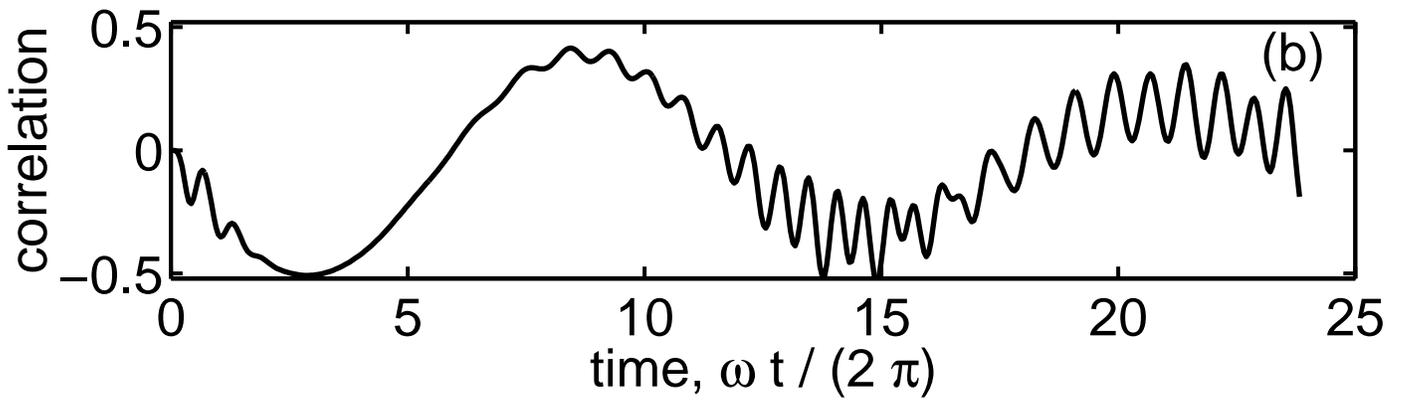
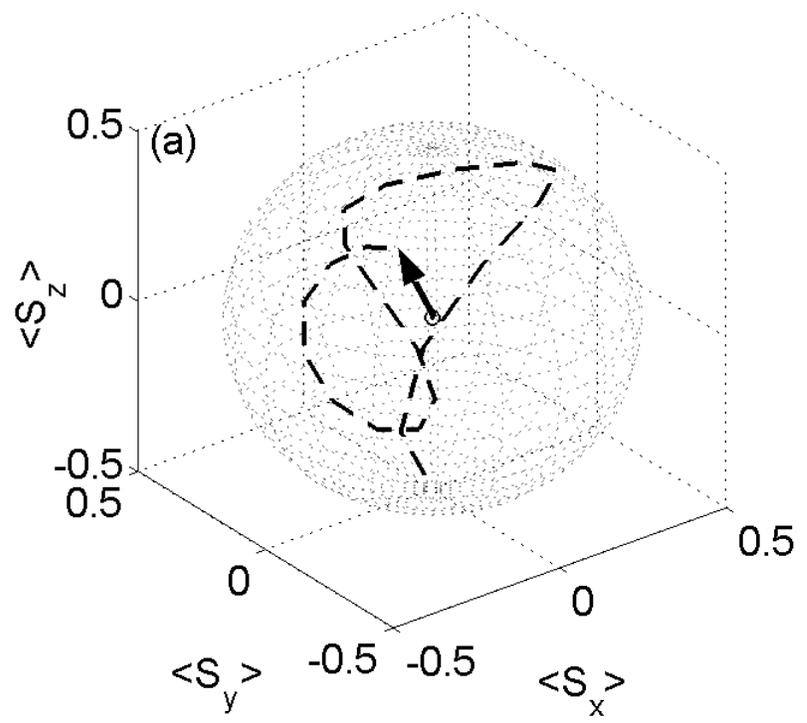
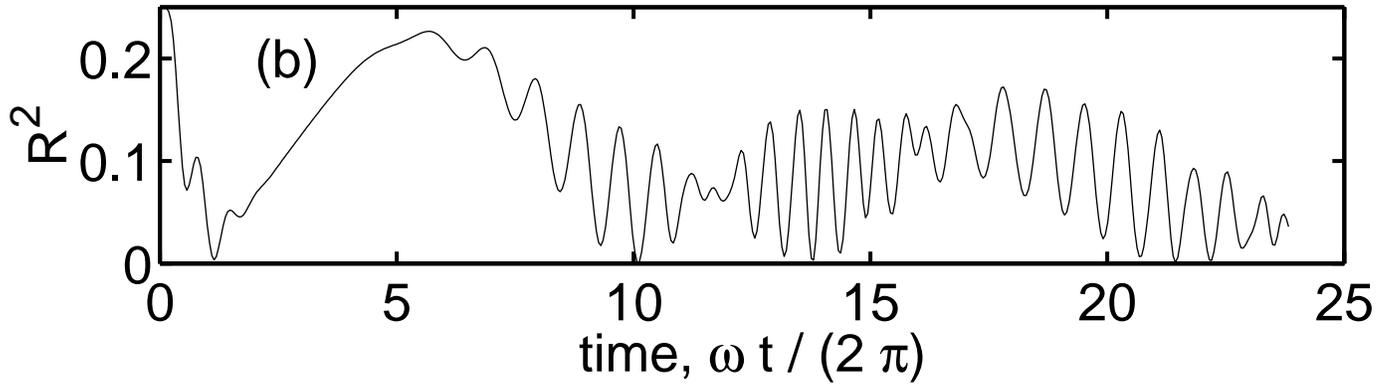
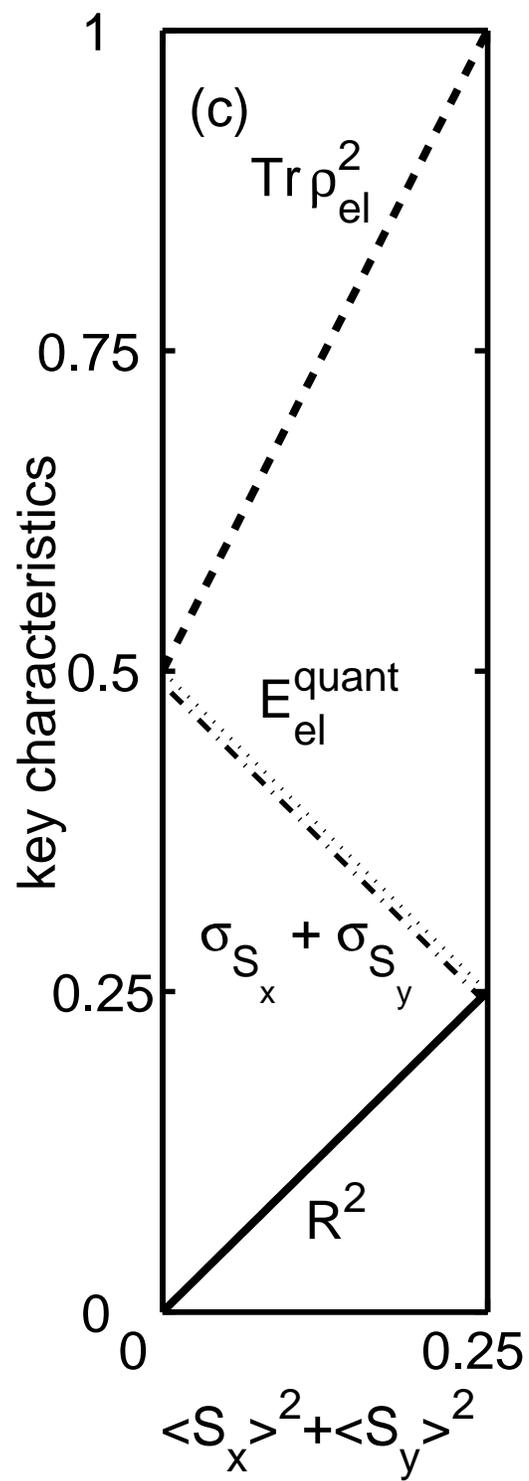
\end{document}